\documentclass{aa}  

\usepackage{graphicx}
\usepackage{txfonts}
\usepackage[colorlinks]{hyperref}
\begin{document}

   \title{Broad-band transmission spectrum and K-band thermal emission of \object{WASP-43b} as 
          observed from the ground\thanks{Based on observations collected with the Gamma 
          Ray Burst Optical and Near-Infrared Detector (GROND) on the MPG/ESO 2.2-meter
          telescope at La Silla Observatory, Chile. Programme 088.A-9016 (PI: Chen)}}
   

   \author{G. Chen\inst{1,2,3}
          \and
           R. van Boekel\inst{3}
          \and
           H. Wang\inst{1}
          \and
           N. Nikolov\inst{3,4}
          \and
           J.~J. Fortney\inst{5}
          \and
           U. Seemann\inst{6}
          \and
           W. Wang\inst{7}
          \and
           L. Mancini\inst{3}
          \and
           Th. Henning\inst{3}
          }
   \institute{Purple Mountain Observatory, \& Key Laboratory for Radio Astronomy, Chinese 
             Academy of Sciences, 2 West Beijing Road, Nanjing 210008, China\\
              \email{guochen@pmo.ac.cn}
         \and
             University of Chinese Academy of Sciences, 19A Yuquan Road, Beijing 100049, China
         \and
             Max Planck Institute for Astronomy, K\"onigstuhl 17, 69117 Heidelberg, Germany
         \and
             Astrophysics Group, University of Exeter, Stocker Road, EX4 4QL, Exeter, UK
         \and
             Department of Astronomy and Astrophysics, University of California, Santa Cruz, CA 95064, USA
         \and
             Institut f\"ur Astrophysik, Friedrich-Hund-Platz 1, 37077 G\"ottingen, Germany
         \and
             Key Laboratory of Optical Astronomy, National Astronomical Observatories, Chinese 
             Academy of Sciences, Beijing 100012, China
             }

   \date{Received 24 September 2013; accepted 8 January 2014}
   \titlerunning{Transit and occultation observations of \object{WASP-43b}}
   \authorrunning{G. Chen et al.}

 
  \abstract
  {} 
   {\object{WASP-43b} is the closest-orbiting hot Jupiter, and has a high bulk 
   density. It causes deep eclipse depths in the system light curve in both 
   transit and occultation attributed to the cool temperature and small radius 
   of its host star. We aim at securing a broad-band transmission spectrum and 
   detecting its near-infrared thermal emission in order to characterise its 
   atmosphere.}
   {We observed one transit and one occultation event simultaneously 
   in the $g'$, $r'$, $i'$, $z'$, $J$, $H$, $K$ bands using the GROND instrument on 
   the MPG/ESO 2.2-meter telescope, in which the telescope was heavily defocussed 
   in staring mode. After modeling the light curves, we derived wavelength-dependent 
   transit depths and flux ratios, and compared them to atmospheric models.}
   {From the transit event, we have independently derived \object{WASP-43}'s system 
   parameters with high precision, and improved the period to be {0.81347437(13)}\,days 
   based on all the available timings. No significant variation in transit depths 
   is detected, { with the largest deviations coming from the $i'$-, $H$-, and $K$-bands. 
   Given the observational uncertainties}, the broad-band transmission spectrum can be explained 
   by either a flat featureless straight line that indicates thick clouds, synthetic 
   spectra with absorption signatures of atomic Na/K or molecular TiO/VO that indicate 
   cloud-free atmosphere, or a Rayleigh scattering profile that indicates high-altitude 
   hazes. From the occultation event, we have detected planetary dayside thermal emission 
   in the $K$-band with a flux ratio of {0.197 $\pm$ 0.042}\%, which confirms previous 
   detections obtained in the 2.09~$\mu$m narrow band and $K_S$-band. The $K$-band 
   brightness temperature {1878 $^{+108}_{-116}$}~K favors an atmosphere with poor day- 
   to night-side heat redistribution. We also have a  marginal detection in the $i'$-band 
   ({0.037 $^{+0.023}_{-0.021}$}\%), corresponding to $T_B$ = {2225 $^{+139}_{-225}$}~K, 
   which is either a false positive, a signature of non-blackbody radiation at this 
   wavelength, or an indication of reflective hazes at high altitude. }
   {}


   \maketitle

\section{Introduction}\label{sec1}

   Transiting hot Jupiters are highly valuable in the characterization study of 
   planetary orbits, structures and atmospheres. They orbit the host stars closely 
   whilst they possess relatively large sizes and masses, thereby producing large 
   transit and radial velocity signals, which together result in precise determination 
   of the planetary system parameters, such as planetary mass, radius, surface gravity, 
   orbital distance, stellar density etc \citep[e.g.][]{2003ApJ...585.1038S,
   2007MNRAS.379L..11S}. Based on these fundamental parameters, the planetary composition 
   and internal structure could be investigated, providing us an insight view of the 
   planetary formation, evolution and migration history \citep[e.g.][]{2005AREPS..33..493G,
   2006A&A...453L..21G,2007prpl.conf..701C,2007ApJ...659.1661F}. 
   
   With high incident irradiation from the host stars, transiting hot Jupiters also 
   play a crucial role in planetary atmospheric characterization. Through observations 
   of primary transits, transmission spectrum can be constructed by measuring the 
   transit depths at different wavelengths, which carries information of planetary 
   atomic (e.g. Na, K) and molecular (e.g. H$_2$O, CH$_4$, CO) absorption features 
   when the stellar lights are transmitted in the planetary day-night terminator 
   region \citep[][etc.]{2000ApJ...537..916S,2008ApJ...678.1419F,2010ApJ...709.1396F}. 
   In a similar way, thermal emission spectrum can be obtained by measuring the 
   occultation depths (approximately the planet-to-star flux ratios) during secondary 
   eclipses (i.e. occultations), which puts constraints on both chemical composition 
   and temperature structure of the dayside atmosphere \citep[][etc.]{1997ApJ...491..856B,
   1999ApJ...512..843B,2006ApJ...650.1140B,2008ApJ...678.1436B,2008ApJ...678.1419F,
   2009ApJ...707...24M}. 
   
   Abundant observations have been performed to characterize the atmospheres for dozens 
   of hot Jupiters in these two aspects \citep[see the review by][]{2010ARA&A..48..631S}. 
   However, very few hot Jupiters have been studied in both aspects which are complementary 
   and could potentially constrain the origin of thermal inversions 
   \citep[e.g.][]{2003ApJ...594.1011H,2007ApJ...668L.171B,2008ApJ...678.1419F,2010ApJ...725..261M}, 
   as the absorbers that cause thermal inversions also imprint their spectral signatures 
   on transmission spectra. This kind of studies have started to emerge on some typical 
   cases of hot Jupiters, for example, \object{HD 189733b} \citep{2013MNRAS.432.2917P}, 
   \object{WASP-12b} \citep{2013Icar..225..432S} and \object{WASP-19b} \citep{2013ApJ...771..108B}.
   
   \object{WASP-43b} is a hot Jupiter orbiting a K7V star every 0.81 days, first discovered 
   by \citet{2011A&A...535L...7H}. It has the smallest orbital distance to its host star 
   among the Jupiter-size planets. The host star is active as indicated by the presence 
   of strong Ca H+K emission. \citet{gillon2012} significantly improved the \object{WASP-43} 
   system parameters based on 20 transits observed with the 60-cm robotic TRAPPIST 
   telescope and 3 transits with the 1.2-m Euler Swiss telescope. The deduced planetary 
   mass (2.034 $\pm$ 0.052\,$M_{\rm{Jup}}$) and radius (1.036 $\pm$ 0.019\,$R_{\rm{Jup}}$) 
   indicate a high bulk density that favors an old age and a massive core. They also 
   observed the occultations of \object{WASP-43b} at narrow bands with VLT/HAWK-I, resulting 
   in a thermal emission detection at 2.09~$\mu$m ($F_{\rm{p}}/F_{\star}=0.156\pm0.014\%$) 
   and a tentative detection at 1.19~$\mu$m (0.079 $\pm$ 0.032\%). \citet{2013arXiv1302.7003B} 
   performed atmospheric modeling based on their Warm {\it Spitzer} detections at 
   3.6~$\mu$m (0.346 $\pm$ 0.012\%) and 4.5~$\mu$m (0.382 $\pm$ 0.015\%) in combination 
   with the two ground-based near infrared (NIR) detections of \citet{gillon2012}, 
   which rules out a strong thermal inversion in the dayside atmosphere, and suggests 
   inefficient day-night energy redistribution. Recently, \citet{2013ApJ...770...70W} 
   carried out occultation observations with the WIRCam instrument on the Canada-France-Hawaii 
   telescope, and detected the thermal emission in the $H$-band (0.103 $\pm$ 0.017\%) 
   and $K_S$-band (0.194 $\pm$ 0.029\%). However, current data are still insufficient 
   to constrain the chemical composition of \object{WASP-43b}'s atmosphere. 
   
   In 2011, we started a project to characterize hot-Jupiter atmospheres using the GROND instrument 
   \citep[e.g. Paper I on \object{WASP-5b};][]{paper1}, which is capable of doing simultaneous multi-band 
   photometry. This technique now has been performed on several transiting planets to investigate their atmospheres 
   \citep{2012A&A...538A..46D,2012MNRAS.422.3099S,2013A&A...551A..11M,2013MNRAS.430.2932M,2013MNRAS.436....2M,
   2013A&A...553A..26N,2013ApJ...770...95F,2013MNRAS.434..661C,2013arXiv1308.6765N}. Among our targets, 
   \object{WASP-43b} is favorable for observation because of its large transit depth $\sim$2.5\% 
   and high incident irradiation $\sim$9.6$\times$10$^8$~erg\,s$^{-1}$\,cm$^{-2}$ \citep{gillon2012}. 
   If zero albedo and zero day to night heat redistribution are assumed, an equilibrium 
   temperature of 1712 K is derived. We aim to construct a broad-band transmission 
   spectrum for \object{WASP-43b} and also to detect its NIR thermal emission. 
   
   This paper is organized as follows: Sect.~\ref{sec2} summarises the transit and 
   occultation observations, as well as the data reduction. Sect.~\ref{sec3} describes 
   the process of transit and occultation light curve modeling, including the re-analysis 
   of light curves obtained by amateur astronomers. Sect.~\ref{sec4} reports the newly 
   derived orbital ephemeris and fundamental physical parameters, and presents discussion 
   on \object{WASP-43b}'s atmospheric properties. Sect.~\ref{sec5} gives the overall 
   conclusions.

\section{Observations and Data Reductions}\label{sec2}
   \begin{table*}
     \caption{Summary of the \object{WASP-43b} Observations with the GROND instrument}
     \label{tab:log}
     \small
     \centering
     \begin{tabular}{ccp{0.5cm}ccp{0.5cm}cccp{0.2cm}ccc}
       \hline\hline
       Date & Start/end time & Type & Airmass & Moon & Filter & $N_{\rm{obs}}$ & $t_{\mathrm{exp}}$ & Aperture 
       & $N_{\rm{ref}}$ & \multicolumn{2}{c}{$\beta$} & $\sigma_{120\mathrm{s}}$\\
       
       (UT) & (UT) & & & illum. ($d$) & & & (sec) & ($''$) & & $\beta_{\chi}$ & $\beta_r$ & (ppm)\\
       \hline
       2012-01-08 & 05:24$\to$08:21 & Tran. & 1.29$\to$1.06$\to$1.07 & 99\% {(65$^{\circ}$)}
                          & $g'$ & 81  & 90 & 3.6,4.9,6.0  & 1 & {1.26} & {1.21} & {678}\\
                  & & & & & $r'$ & 81  & 90 & 5.5,8.4,9.0  & 1 & {1.40} & 1.00 & 504\\
        	          & & & & & $i'$ & 81  & 90 & 6.2,9.0,9.6  & 1 & {1.23} & 1.00  & {645}\\
        	          & & & & & $z'$ & 81  & 90 & 4.7,7.6,8.4  & 1 & {1.57} & 1.00  & {798}\\
        	          & & & & & $J$  & 244 & 3$\times$8 & {6.3,11.1,15.6}  & 3 & {0.59} & {1.97}  & {1241}\\
        	          & & & & & $H$  & 244 & 3$\times$8 & {6.3,11.7,15.6}  & 1 & {1.01} & {2.83}  & {1269}\\
        	          & & & & & $K$  & 244 & 3$\times$8 & {5.1,9.9,14.4}  & 2 & {0.64} & {1.50}  & {1269}\\
       \hline
       2012-03-03 & 02:52$\to$05:59& Occ. & 1.12$\to$1.06$\to$1.18 & 69\% {(63$^{\circ}$)} 
                          & $g'$ & 98  & 90 & 5.4,7.3,8.1  & 2 & {1.57} & 1.00 & {673}\\
                  & & & & & $r'$ & 98  & 90 & 6.5,9.0,9.6  & 1 & {2.31} & {1.52} & {475}\\
       	          & & & & & $i'$ & 98  & 90 & 5.5,7.7,8.5  & 2 & {2.58} & 1.00 & {604}\\
       	          & & & & & $z'$ & 98  & 90 & 5.5,7.7,8.5  & 1 & {2.50} & {1.49} & {575}\\
       	          & & & & & $J$  & 389 & 4.5$\times$4 & {8.4,11.4,15.6}  & 3 & {1.30} & {2.53} & {953}\\
       	          & & & & & $H$  & 389 & 4.5$\times$4 & {6.9,11.1,15.6}  & 3 & {0.95} & {3.38} & {885}\\
       	          & & & & & $K$  & 389 & 4.5$\times$4 & {6.0,9.6,14.4}  & 2 & {0.47} & {1.09} & {1425}\\
       \hline
     \end{tabular}
     \tablefoot{\small
       Type "Tran."/"Occ." corresponds to transit/occultation, respectively. "Moon illum." is the fraction 
       of the Moon that is illuminated{, while the bracketed $d$ is the minimum distance to the Moon}. 
       $N_{\rm{obs}}$ is the number of retrieved images, while $N_{\rm{ref}}$ 
       is the number of reference stars to create a composite reference light curve. $t_{\mathrm{exp}}$ 
       refers to integration time for $g'r'i'z'$, and detector integration time (DIT) times the number of DIT 
       ($N_{\rm{DIT}}$) for $JHK$. Aperture sizes refer to the optimal aperture, inner/outer annuli radii 
       adopted in the photometry. { $\beta_{\chi}$ and $\beta_r$ are the $\chi_{\nu}$-rescaling and red 
       noise rescaling factors as described in Sect.~\ref{sec301}. }
       $\sigma_{120\mathrm{s}}$ refers to the standard deviation of light-curve O--C residuals binned in every 
       2 minutes. { The last three columns are calculated from the global joint 
       analysis with wavelength-dependent radius (or flux ratios), i.e. Method 3 as described in Sect.~\ref{sec302}.}
       }
   \end{table*}
   
   We observed one primary transit and one secondary eclipse events of \object{WASP-43b} 
   with the GROND instrument mounted on the MPG/ESO 2.2-meter telescope at La Silla in Chile. 
   GROND is an imaging instrument primarily designed for the investigation of gamma-ray burst 
   afterglows and other transients simultaneously in seven bands: Sloan $g'$, 
   $r'$, $i'$, $z'$ and near-infrared $J$, $H$, $K$ \citep{2008PASP..120..405G}. The light 
   goes through dichroics, and is split into optical and NIR arms and further into seven bands.
   The optical arm employs backside illuminated $2048\times2048$ E2V CCDs without 
   anti-blooming structures, which have a field of view (FOV) of 5.4\,$\times$\,5.4\,arcmin$^2$ 
   and a pixel scale of $0\farcs158$, and store data in FITS file with four extensions. 
   The NIR arm employs $1024\times1024$ Rockwell HAWAII-1 arrays, with an FOV of 
   10\,$\times$\,10\,arcmin$^2$ and a pixel scale of 0\farcs60, which place data of three 
   bands side by side in a single FITS file. The GROND guide camera is placed outside 
   of the main GROND vessel, resulting in a guiding FOV which is $23'$ south of the scientific 
   imaging FOV. Taking into account the different FOV scales between optical and red arms, 
   the orientation of guiding system, and the inclusion of as many reference stars as possible, 
   a compromise between these factors is necessary when designing the observing strategy. 
   Nevertheless, GROND is still a potentially good instrument for exoplanet observations, 
   since it has very few moving parts and the capability of multi-band observations with large 
   wavelength coverage.
   
   The observing strategies for both transit and occultation were the same. We heavily 
   defocussed the telescope during both observations in order to spread the light of 
   stars onto more pixels, which as a result could reduce the noise arising from small 
   number of pixels and avoid reaching the non-linear regime. When achievable, we 
   did sky measurements both before and after the science time series, which were designed 
   in a 20-position dither pattern around the scientific FOV. According to the experience 
   that we had in the observations of \object{WASP-5b} \citep[Paper I]{paper1}, staring 
   mode is more stable and introduces less instrumental systematic effects than nodding 
   mode. Therefore, for the \object{WASP-43b} observations, we made the telescope staring on 
   the target during the 3-hour long science time series. The observations were carried out 
   in the $g'r'i'z'JHK$ filters simultaneously. The final observing duty cycle is determined 
   by the compromise between the integration times of optical arm and NIR arm, as they 
   are not fully independent in operation. The summary of both transit and occultation 
   observations are listed in Table~\ref{tab:log}.
   
  \subsection{Transit observation}\label{sec201}
 
\begin{figure*}
     \centering
     \includegraphics[width=0.32\hsize]{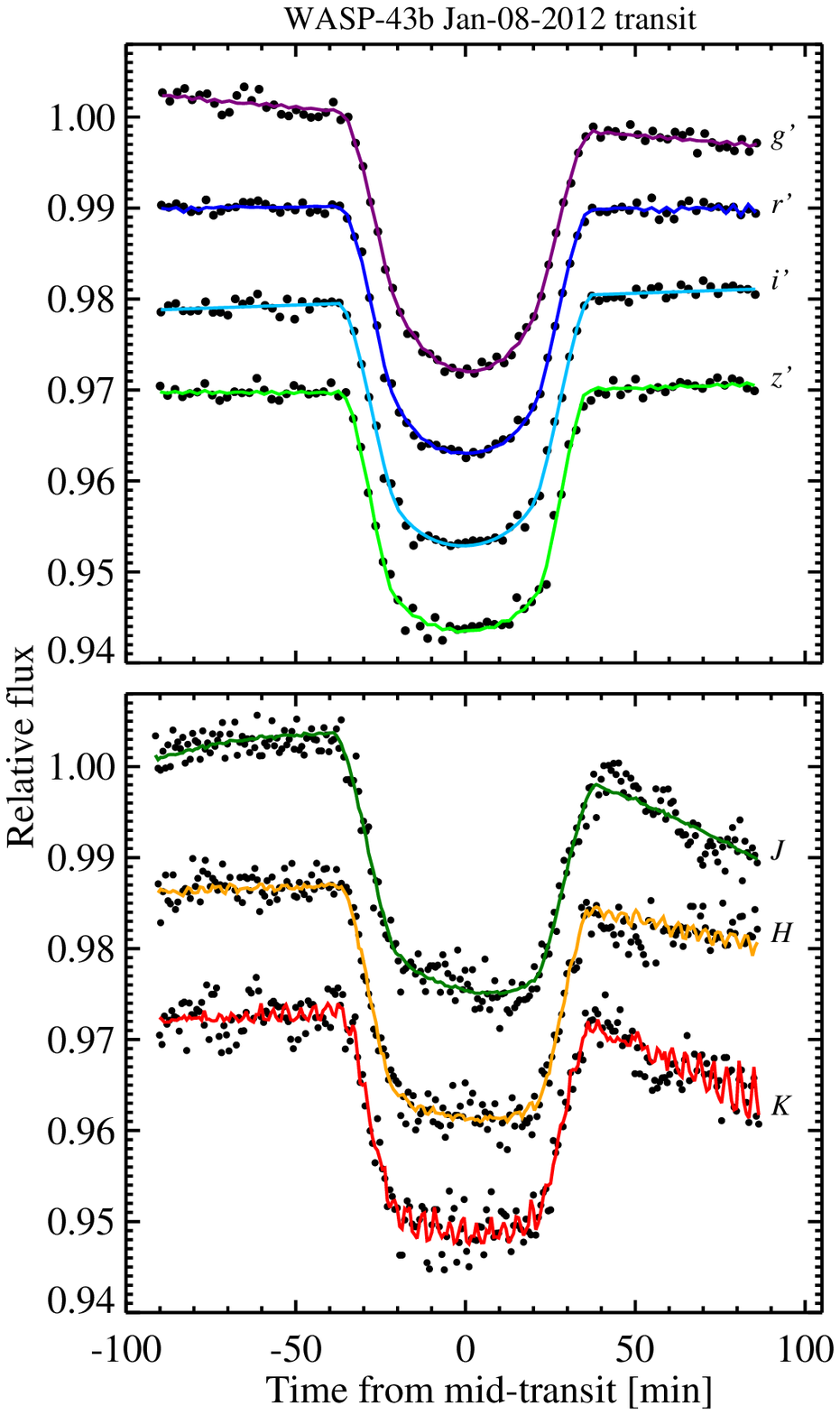}
     \includegraphics[width=0.32\hsize]{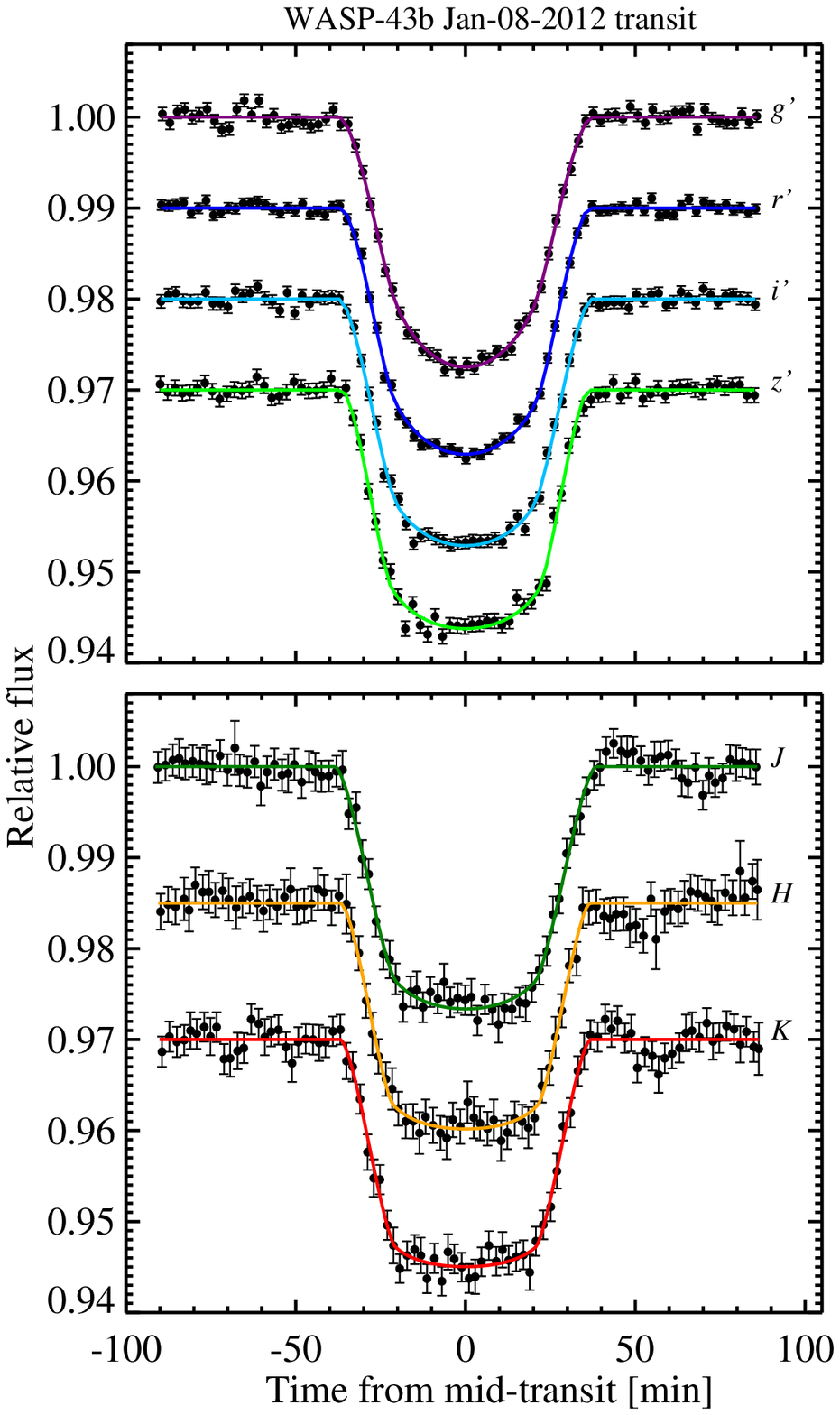}
     \includegraphics[width=0.32\hsize]{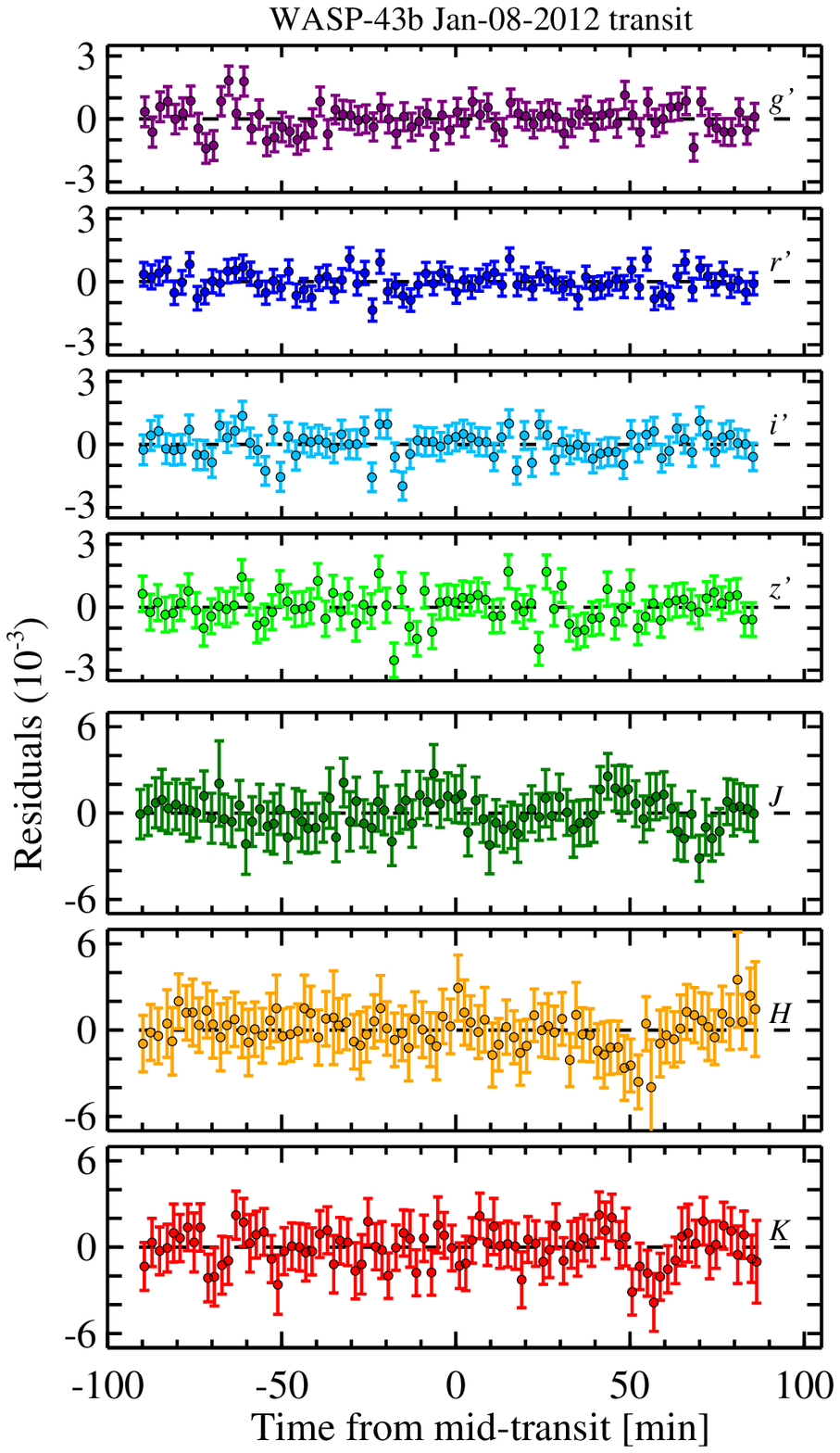}
     \caption{Primary transit light curves of \object{WASP-43b} as observed with the GROND 
              instrument mounted on the ESO/MPG 2.2\,m telescope. In each panel, shown from 
              top to bottom are transit light curves of $g'r'i'z'JHK$. {\it Left panel} shows 
              light curves that are normalized by the reference stars. {\it Middle panel} 
              shows baseline-corrected light curves which are binned in every 2 minutes for 
              display purpose (see Sect.~\ref{sec301} for baseline correction). {\it Right panels} 
              show the best-fit light-curve residuals, also binned per 2 minute. Best-fit 
              models for all panels are overlaid in solid or dashed lines.}
     \label{fig:tran}
   \end{figure*}

   The transit event was observed on January 8 2012, from 05:24 to 08:21 UT, which 
   covered $\sim$50 minutes both before expected ingress and after expected egress 
   (see Fig.~\ref{fig:tran}). This night was of high relative humidity, ranging from 
   40\% to 60\%. The moon was illuminated by 99\%, with a minimum distance of 65$^{\circ}$ 
   to WASP-43. The airmass was well below 1.30. Although the telescope was heavily 
   defocussed, the radius of the donut-shaped point spread function (PSF) ring varied 
   by as much as $1\farcs0$ during the observation, indicating that the seeing was 
   not very stable and had evident impact on our PSF. The observation was in a good 
   auto-guiding status thanks to a bright star in the guiding FOV. We did sky measurements 
   only before the science time series as the morning twilight had begun at the end 
   of the observation. For the science time series, the optical bands were integrated 
   for 90 seconds in fast read-out mode simultaneously, and 81 frames were recorded, 
   resulting in a duty cycle of $\sim$69\%. The NIR bands were integrated with 8 
   sub-integrations of 3 seconds, which were averaged together before readout. 244 
   frames were recorded in the NIR, resulting in a duty cycle of $\sim$56\%. 
   
  \subsection{Occultation observation}\label{sec202}
 
   The occultation event was observed on March 3 2012, from 02:52 to 05:59 UT, 
   which covered $\sim$70 minutes before expected ingress and $\sim$40 minutes 
   after expected egress (see Fig.~\ref{fig:occ}). Relative humidity for this 
   night was also high ($\sim$45\%), and the moon was illuminated by 69\%, with 
   a minimum distance of 63$^{\circ}$ to WASP-43. The airmass was below 1.20 for 
   the whole event. The resulting PSF donut ring size was stable out-of-eclipse, 
   while it varied a lot during the eclipse, which was caused by the sudden 
   jumps of poor seeing. We did sky measurements both before and after the 
   science observation. The optical bands were integrated with 90 seconds 
   in fast read-out mode, while the NIR bands were integrated with 4 
   sub-integrations of 4.5 seconds. In the end, we collected 98 frames 
   for the optical and 389 frames for the NIR, translating into a duty 
   cycle of 78\% and 63\%, respectively.

  \subsection{Data reduction}\label{sec203}
   Both sets of observations were reduced in a standard way with our IDL\footnote{IDL 
   is an acronym for Interactive Data Language, for details please refer to 
   http://www.exelisvis.com/idl/} codes, which largely make use of NASA IDL Astronomy 
   User's Library\footnote{See http://idlastro.gsfc.nasa.gov/}. Calibration for the 
   optical data includes bias subtraction and flat division. The master frames for both 
   were created from median combination of individual measurements. The twilight sky 
   flat measurements were star-masked and normalized before median combination. 
   
   For the NIR data, the calibration is a little more complicated due to the presence 
   of an electronic odd-even readout pattern along the X-axis. { The calibration 
   includes dark subtraction, readout pattern removal, and flat division}. The 
   master frames for dark, twilight sky flat and sky measurements were created through 
   median combination as well, except that the sigma-clipping filter was applied if 
   necessary.  After the dark was corrected, each image was smoothed with a median 
   filter and compared to the unsmoothed one. The amplitudes of the readout pattern 
   were determined by differentiating median value of each column to the overall median 
   level, and were then corrected in the original dark-subtracted images. The flat field 
   was corrected after the removal of this pattern. 
   
   { According to our experience on other targets \citep[e.g.][]{paper1}, sky 
   subtraction might result in light curves of slightly better precision. In this 
   approach, a sky model is created and subtracted for each individual science frame 
   by optimally combining the scaled before- and after-science sky measurements. 
   This technique is not capable of removing temporal variations, but is helpful for 
   removing spatial variations in principle. However, it is not the case for this 
   study. We found no improvement in the data set by applying the sky subtraction 
   technique. Therefore, we decided to perform our analysis on the data set without 
   sky subtraction. }
   
   After these calibrations, we employed the IDL/DAOPHOT package to perform aperture 
   photometry on \object{WASP-43} as well as nearby comparison stars of similar brightness. 
   The location of each star was determined by IDL/FIND, which calculates the centroid 
   by fitting Gaussians to the marginal x, y distributions. This finding algorithm was 
   applied on the Gaussian-filtered images that were of lower resolution, in which the 
   irregular donut-shaped PSFs became nearly Gaussian after proper convolution. We 
   recorded the FWHM of each star by masking out the central region of the original 
   donut and fitting a Gaussian to the donut wing. Time series of each star was 
   self-normalized with its out-of-transit flux level. We then carefully chose the best 
   comparison star ensemble in each band to correct the first order atmospheric effect 
   on the \object{WASP-43} time series, in the following way: various combination of 
   comparison stars were experimented. The ensemble which leaves \object{WASP-43} with 
   the least light-curve O--C residuals was chosen. In order to find the optimal photometry, 
   40 apertures were laid on the optical images in step of 1 pixel, each with 10 sky annulus 
   sizes in steps of 2 pixels, while 30 apertures were laid on the NIR images in steps of 
   0.5 pixel, each with 10 sky annulus sizes in step of 1 pixel. This resulted in 400 and 
   300 datasets with different aperture settings for the optical and NIR, respectively, of 
   which the one that leaves \object{WASP-43} with the least light-curve O--C scatter was 
   adopted. Table~\ref{tab:log} lists the number of reference stars in use and the final 
   apertures for photometry.
   
   Finally, we extracted the observation time of each frame. In this process, we 
   made the time of each frame centered based on its actual total integration. This 
   UTC time was then converted to Barycentric Julian Date in the Barycentric Dynamical 
   Time standard (BJD$_{\mathrm{TDB}}$) using the IDL procedure written by 
   \citet{2010PASP..122..935E}.

\section{Light curve analysis}\label{sec3}
  \subsection{Light curve modeling}\label{sec301}
   \begin{table}
     \caption{Quadratic limb-darkening coefficients adopted in this work}
     \label{tab:ld}
     \small
     \centering
     \begin{tabular}{ccc}
       \hline\hline
       Filter & $u_1$ & $u_2$\\
       \hline
       $g'$ & 0.867 $\pm$ 0.022 & $-$0.042 $\pm$ 0.020\\
       $r'$ & 0.628 $\pm$ 0.017 & 0.120 $\pm$ 0.011\\
       $i'$ & 0.486 $\pm$ 0.010 & 0.166 $\pm$ 0.005\\
       $z'$ & 0.398 $\pm$ 0.008 & 0.184 $\pm$ 0.004\\
       $J$  & 0.286 $\pm$ 0.009 & 0.229 $\pm$ 0.003\\
       $H$  & 0.135 $\pm$ 0.005 & 0.330 $\pm$ 0.003\\
       $K$  & 0.120 $\pm$ 0.003 & 0.275 $\pm$ 0.003\\
       \hline
     \end{tabular}
   \end{table}

   \begin{figure*}
     \centering
     \includegraphics[width=0.32\hsize]{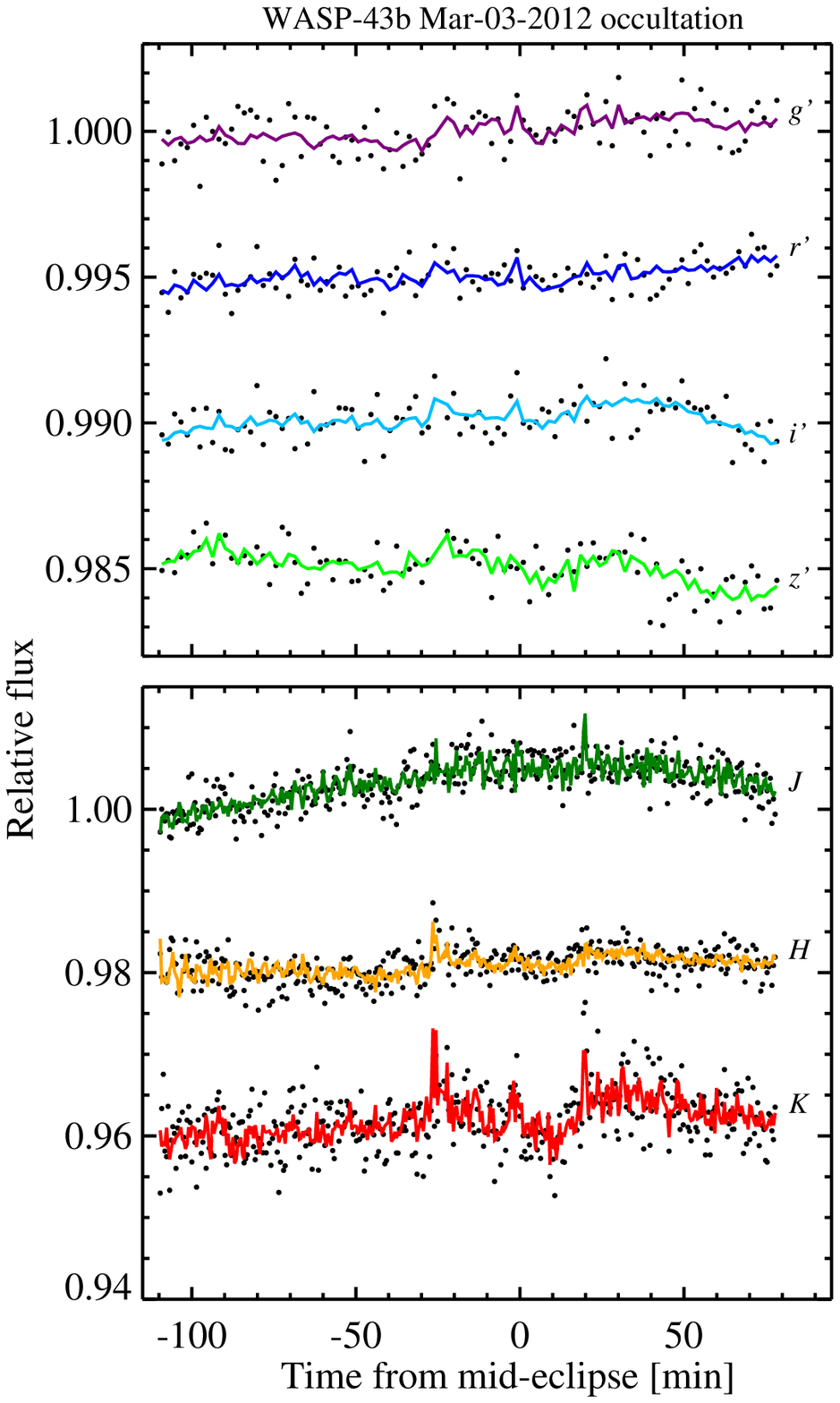}
     \includegraphics[width=0.32\hsize]{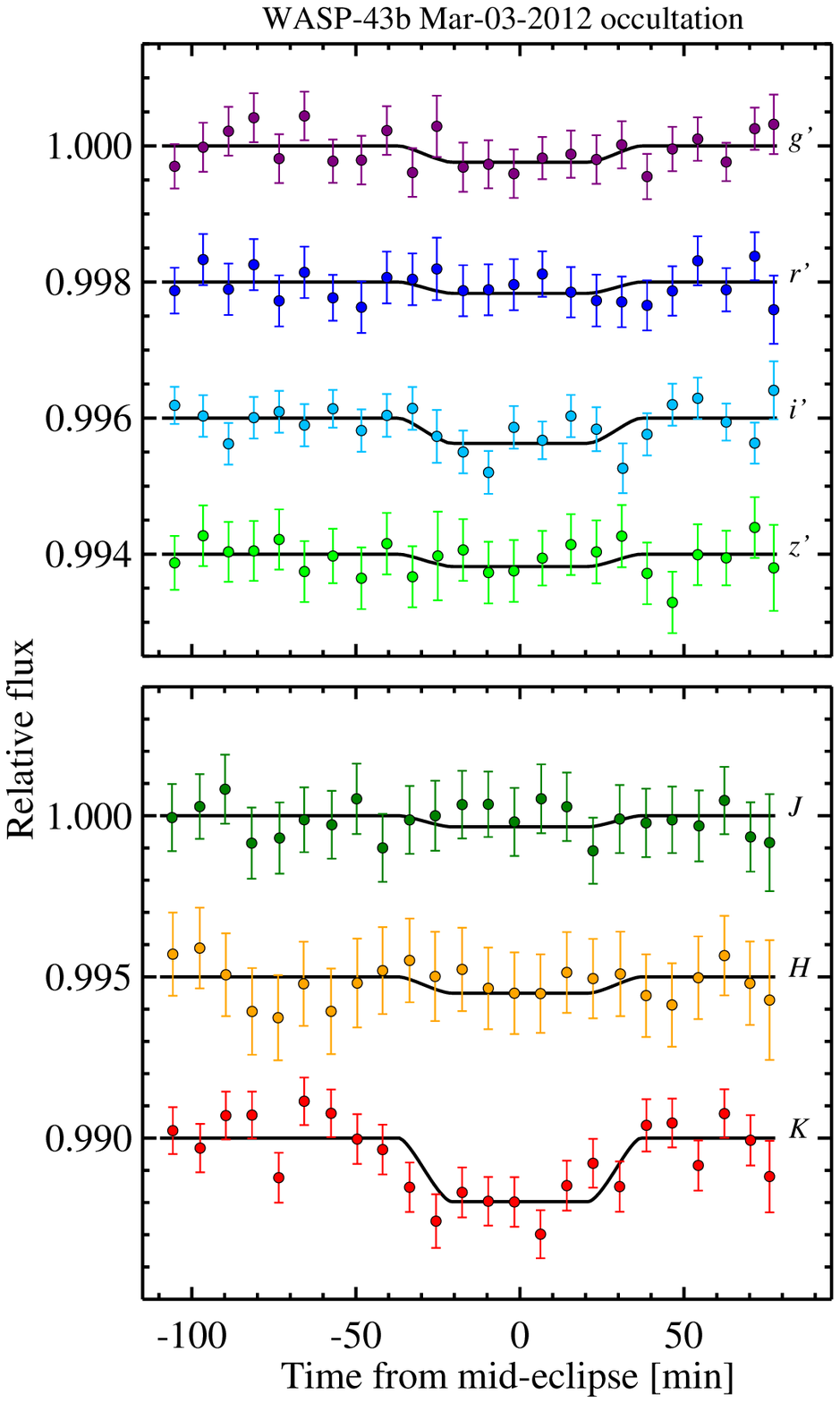}
     \includegraphics[width=0.32\hsize]{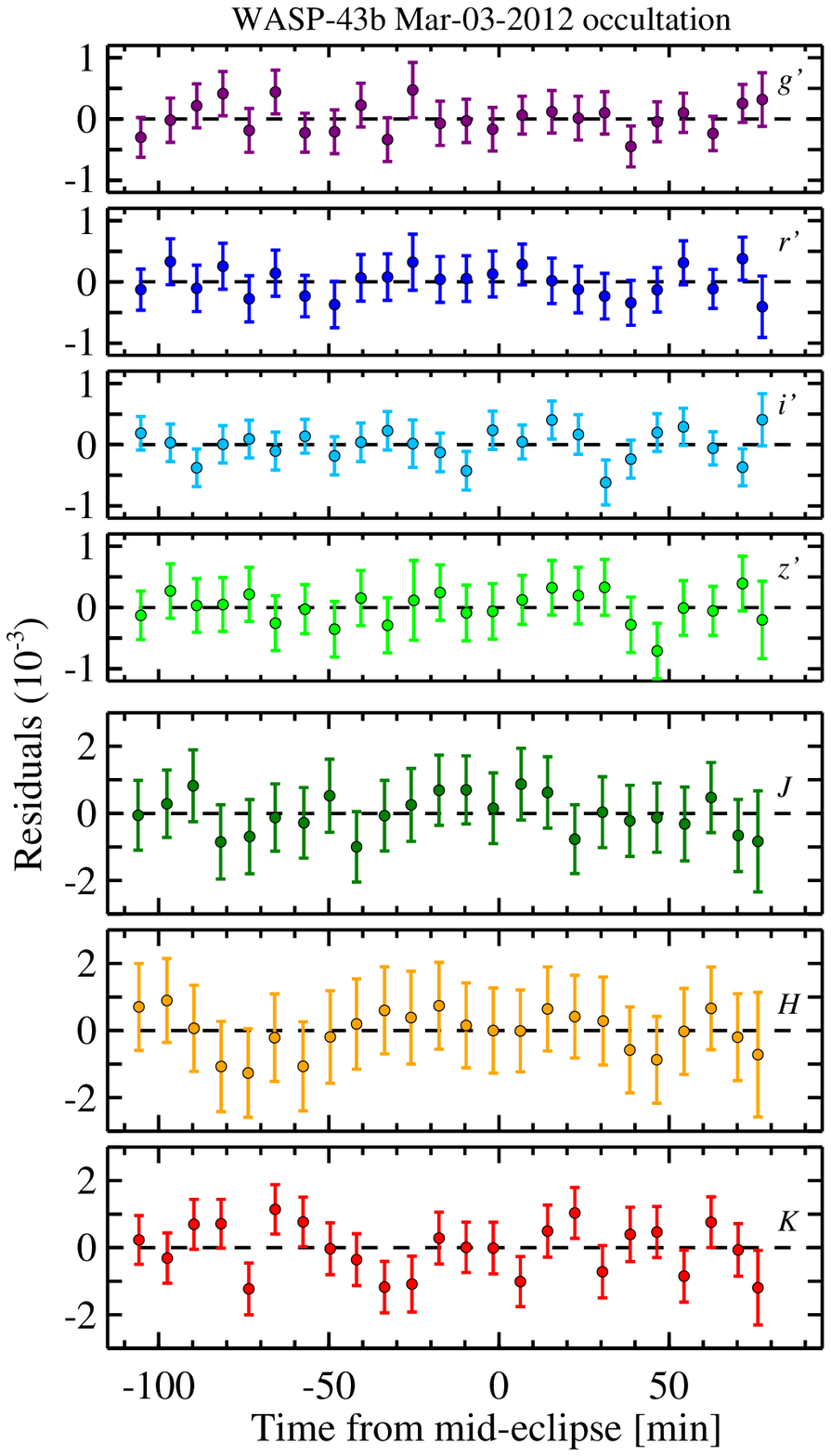}
     \caption{Occultation light curves of \object{WASP-43b} as observed with GROND. 
              In each panel, shown from top to bottom are occultation light curves 
              of $g'r'i'z'JHK$. {\it Left panel} shows light curves that are normalized 
              by the reference stars. {\it Middle panel} shows baseline-corrected 
              light curves which are binned in every 8 minutes for display purpose 
              (see Sect.~\ref{sec301} for baseline correction). {\it Right panels} 
              show the best-fit light-curve residuals, also binned per 8 minute. 
              Best-fit models for all panels are overlaid in solid or dashed lines.
              The occultation depth is detected in the $K$-band ({0.197 $\pm$ 0.042\%}), 
              and marginally detected in the $i'$-band ({0.037 $^{+0.023}_{-0.021}$\%}).}
     \label{fig:occ}
   \end{figure*}

   As shown in the left panels of Fig.~\ref{fig:tran} and \ref{fig:occ}, the reference-corrected 
   light curves still exhibit systematics which are correlated with instrumental parameters 
   or atmospheric conditions (e.g. star's location on the detector, seeing etc). In order to 
   model these light curves properly, we chose to fit the data with models composed of two 
   components: 
   \begin{equation}
      F(\mathrm{mod})=E(p_i)B(x,y,z,s,t)
   \end{equation}
   
   The first component is the analytic light curve model \citep{2002ApJ...580L.171M}, 
   representing the transit or occultation signal itself. For the transit event, a quadratic 
   limb darkening law is adopted:
   \begin{equation}
      I_{\mu}/I_1=1-u_1(1-\mu)-u_2(1-\mu)^2
   \end{equation}
   where $I$ is the intensity and $\mu=\cos\theta$ ($\theta$ is the angle between the emergent 
   intensity and the line of sight). We calculate the two theoretical coefficients $u_1$ and 
   $u_2$ by bilinearly interpolating in the \citet{2011A&A...529A..75C} table, in which 
   the effective temperature $T_{\mathrm{eff}}=4520\pm120$\,K, the surface gravity 
   $\log g_{\star}=4.645\pm0.011$, the metallicity [Fe/H]$=-0.01\pm0.12$ are taken from 
   \citet{gillon2012} while the microturbulence $\xi_r=0.5\pm0.3$~km~s$^{-1}$ from 
   \citet{2011A&A...535L...7H}. Resulting values with uncertainties are set as Gaussian 
   priors (GP; see Equation~\ref{eqn:chi2}) in the subsequent fitting process, and listed 
   in Table~\ref{tab:ld}. Thus this theoretical model $E(p_i)$ contains six free parameters: 
   the orbital inclination $i$, the planet-to-star radius ratio $R_{\rm{p}}/R_{\star}$, the 
   scaled semi-major axis $a/R_{\star}$, the mid-transit point $T_{\rm{mid}}$, the limb-darkening 
   coefficients $u_1$ and $u_2$, while the period $P$ is fixed at literature value (or updated 
   ephemeris iteratively, see Sect.~\ref{sec401}). The eccentricity $e$ is fixed at zero since 
   it cannot be determined by a single transit. 
   
   For the occultation event, $E(p_i)$ becomes: 
   \begin{equation}
   E(T_{\rm{occ}},F_{\rm{p}}/F_{\star})=1-\frac{\lambda_e}{1+(F_{\rm{p}}/F_{\star})^{-1}}
   \end{equation}
   where $\lambda_e$ refers to the Equation 1 in \citet{2002ApJ...580L.171M} in the case of 
   uniform source without limb darkening. This model has two free parameters: the mid-occultation 
   time $T_{\rm{occ}}$ and the planet-to-star flux ratio $F_{\rm{p}}/F_{\star}$, while other 
   parameters are fixed at values inherited from the transit event.
   
   The second component is the baseline correction function, which is used to correct 
   the instrumental and atmospheric systematics, including the effects of the star's location 
   on the detector ($x$, $y$), PSF donut ring size $s$, airmass $z$, 
   and time sequence $t$. We search for the best-fit parameters by minimizing the chi-square:
   \begin{equation}
      \chi^2=\sum\limits_{i=1}^{N}\frac{[F_i(\mathrm{obs}) - F_i(\mathrm{mod})]^2}{\sigma_{F,i}^2(\mathrm{obs})}+\mathrm{GP}\label{eqn:chi2}
   \end{equation}
   We select the best baseline model for each band based on the Bayesian Information 
   Criterions \citep[BICs,][]{Schwarz1978}: $BIC=\chi^2+k\log(N)$, where $k$ is the 
   number of free parameters and $N$ is the number of data points. The chosen baselines 
   with their derived coefficients for both nights are listed in Appendix~\ref{sec:ap01}.
   
   To determine the probability distribution function (PDF) for each parameter, we employ 
   the Markov Chain Monte Carlo (MCMC) technique utilizing the Metropolis-Hastings algorithm 
   with Gibbs sampling \citep[see e.g.][]{2005AJ....129.1706F,2006ApJ...642..505F}. Instead 
   of perturbing the baseline coefficients, we follow the approach of \citet{2010A&A...511A...3G}, 
   in which these coefficients are solved by the Singular Value Decomposition algorithm 
   \citep[SVD,][]{press1992}. At each MCMC step, a jump parameter is randomly picked out, 
   and the analytic light curve model $E(p_i)$ is divided from the light curve. The baseline 
   coefficients are then determined by linear least square minimization using SVD. If the 
   resultant $\chi^2$ is less than the previous $\chi^2$, this jump is directly accepted. 
   However, if it's larger, we still accept it with a probability of $\exp{(-\Delta\chi^2/2)}$. 
   Before a chain starts, we optimize the step scale using the method proposed by 
   \citet{2006ApJ...642..505F} so that the acceptance rate is $\sim$0.44. After several chains 
   are completed, we do \citet{gelman1992} statistics to check whether they are well mixed and 
   converged. In the end, we discard the first 10\% links of each chain and calculate the median 
   value, and the 15.865\% and 84.135\% level of the marginalized distribution of the remaining 
   links, which are recorded as the best-fit parameters and 1-$\sigma$ lower and upper uncertainties.
   
\begin{figure}
     \centering
     \includegraphics[width=\hsize]{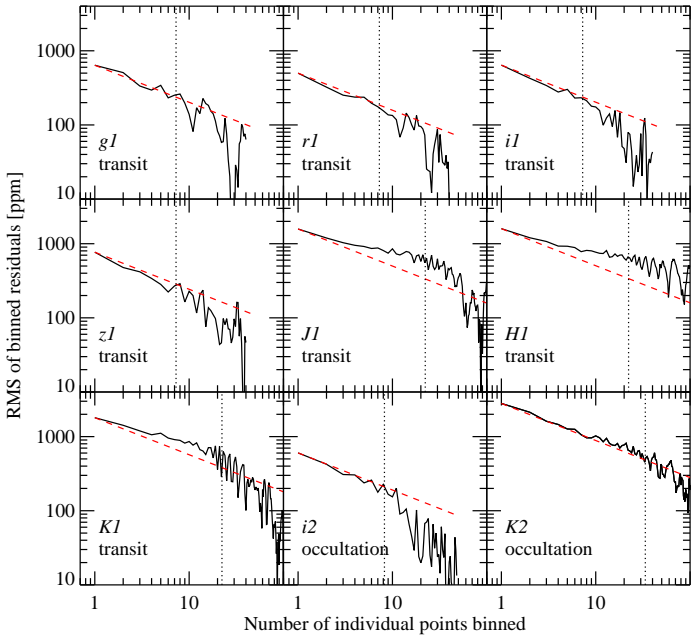}
     \caption{Standard deviation of light-curve O--C residuals binned in different time resolutions, 
              showing the level of time-correlated noise. Red dashed line indicates the expected 
              Poisson-like noise, i.e. standard deviation of unbinned O--C residuals over square-root 
              of $N$. Vertical dotted line shows corresponding ingress/egress duration. The number 
              following filter name indicates either primary transit (1) or occultation (2).}
     \label{fig:rmsbin}
   \end{figure}

   Since the photometric uncertainties sometimes do not reflect the actual properties 
   of the light curve data points, either due to under-/over-estimation of noise or due to 
   time-correlated noise, we performed rescaling on the uncertainties in two steps. For 
   each MCMC fitting process, we always initially ran several chains of 10$^5$ links on 
   the data with original photometric uncertainties. In the first step of rescaling, we 
   calculated the reduced $\chi_{\nu}^2$ for the best-fit model, and recorded the first 
   rescaling factor $\beta_{\chi}=\sqrt{\chi_{\nu}^2}$ in order to force the best fit to 
   have a reduced chi-square equal to unity. In the second step, we followed the approach
   of \citet{2008ApJ...683.1076W} to account for the effect of time-correlated red noise, 
   in which two methods were utilized. In the "time-averaging" (TA) method 
   \citep[see e.g.][]{2006MNRAS.373..231P}, red noise remains unchanged while white noise 
   scales down with larger binning (see Fig.~\ref{fig:rmsbin}). Standard deviations for 
   the best-fit residuals without binning and with binning in different time resolutions, 
   ranging from 10 minutes up to the duration of ingress/egress, were calculated to estimate 
   this factor:
   \begin{equation}
     \beta_{\mathrm{TA},N}=\frac{\sigma_N}{\sigma_1}\sqrt{\frac{N(M-1)}{M}}
   \end{equation}
   where $\sigma_N$ is the standard deviation of residuals binned in every $N$ points, 
   and $M$ is the number of bins. The final $\beta_{\mathrm{TA}}$ is taken to be the median 
   value of several largest $\beta_{\mathrm{TA},N}$, of which outliers could be removed 
   while the information of red noise component remains unchanged. In the "prayer-bead" (PB) 
   method \citep[see e.g.][]{2008MNRAS.386.1644S}, the shape of time-correlated noise was 
   preserved. The best-fit residuals were shifted cyclicly from $i$th to $i$+1th positions, 
   and added back to the best-fit model, while off-position data points at the end were 
   wrapped back to the beginning. This was repeated in an opposite direction as well, so 
   that $2N-1$ synthetic light curves were created and modeled. The ratio between uncertainties 
   derived from this process and original MCMC process was recorded as $\beta_{\mathrm{PB}}$. 
   Finally, we adopted the red noise factor as $\beta_r=\max(\beta_{\mathrm{TA}},\beta_{\mathrm{PB}},1)$. 
   We then rescaled the original uncertainties by $\beta_{\chi}\times\beta_r$ {(see 
   corresponding entries in Table~\ref{tab:log} and \ref{tab:lcpar})}, and ran 
   several more chains of 10$^6$ links to find the final best-fit parameters.

  \subsection{Fitting of transit light curves}\label{sec302}
   \begin{figure}
     \centering
     \includegraphics[width=\hsize]{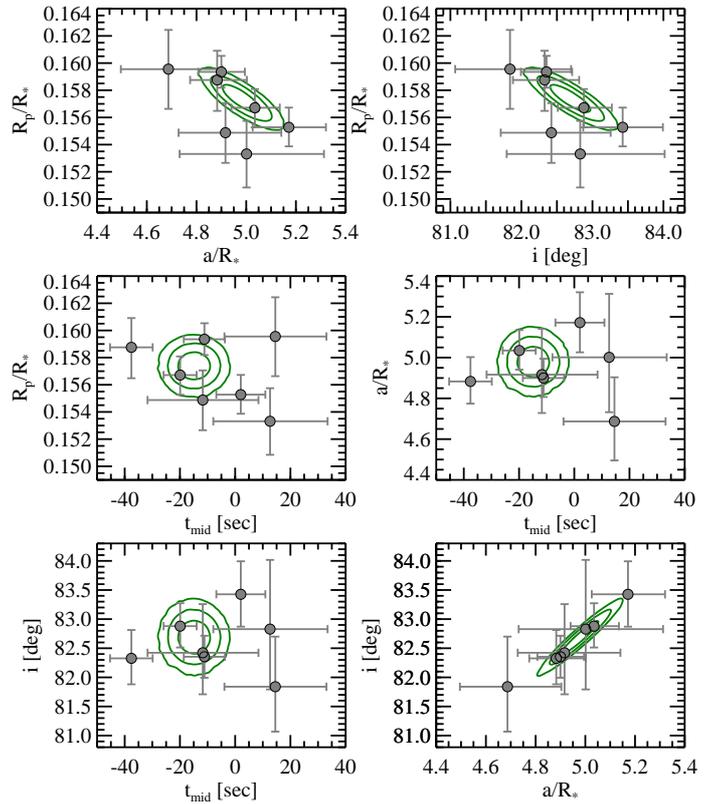}
     \caption{Correlations between pairs of jump parameters for the transit light curves. 
              Black circles with error bars are the results from individual analysis (Method 
              1) on the seven-band light curves, while contours indicate the joint probability 
              distribution function (PDF) from global joint analysis with common radius 
              (Method 2). Both methods have consistent correlation. }
     \label{fig:tranpdf}
   \end{figure}
  
   \begin{table*}
     \caption{Results of the analysis on the Jan-08-2012 transit light curves}
     \label{tab:lcpar}
     \small
     \centering
     \begin{tabular}{cccclcrccc}
       \hline\hline
       Filter & $i$          & $a/R_{\star}$ & \multicolumn{2}{c}{$R_{\rm{p}}/R_{\star}$} 
         & $T_{\mathrm{mid}}$\tablefootmark{b} & \multicolumn{1}{c}{O--C\tablefootmark{c}} 
         & \multicolumn{2}{c}{$\beta$\tablefootmark{d}} & $\sigma_{120\mathrm{s}}$\tablefootmark{d}\\\cline{4-5\smallskip}
       
              & ($^{\circ}$) &         & indi. analysis & joint analysis\tablefootmark{a} 
         & (BJD$_{\mathrm{TDB}}$$-$2450000) & \multicolumn{1}{c}{(min)} & $\beta_{\chi}$ & $\beta_r$ & (ppm)\\
       \hline
      \multicolumn{9}{c}{{\it This work: Method 1 -- individual analysis}}\dotfill\\\smallskip
       $g'$ & {82.33 $^{+0.49}_{-0.45}$} & {4.88 $^{+0.12}_{-0.11}$} & {0.1588 $^{+0.0022}_{-0.0023}$} & 
          {0.15750 $^{+0.00108}_{-0.00109}$} & {5934.791927 $^{+0.000089}_{-0.000089}$} & {-0.45 $^{+0.13}_{-0.13}$} & {1.18} & {1.00} & 639\\\smallskip
       $r'$ & {82.88 $^{+0.39}_{-0.37}$} & {5.03 $^{+0.10}_{-0.09}$} & {0.1567 $^{+0.0014}_{-0.0014}$} & 
          {0.15741 $^{+0.00083}_{-0.00083}$} & {5934.792132 $^{+0.000069}_{-0.000069}$} & {-0.15 $^{+0.10}_{-0.10}$} & {1.38} & 1.00 & 499\\\smallskip
       $i'$ & {82.35 $^{+0.36}_{-0.36}$} & {4.90 $^{+0.10}_{-0.09}$} & 0.1594 $^{+0.0012}_{-0.0012}$ & 
          {0.15847 $^{+0.00077}_{-0.00078}$} & {5934.792233 $^{+0.000086}_{-0.000087}$} & {-0.01 $^{+0.12}_{-0.12}$} & {1.22} & 1.00 & 641\\\smallskip
       $z'$ & {83.43 $^{+0.57}_{-0.56}$} & {5.17 $^{+0.15}_{-0.15}$} & 0.1553 $^{+0.0014}_{-0.0014}$ & 
          {0.15720 $^{+0.00088}_{-0.00089}$} & {5934.792386 $^{+0.000103}_{-0.000103}$} & {0.21 $^{+0.15}_{-0.15}$}  & {1.52} & 1.00 & 768\\\smallskip
       $J$  & {81.84 $^{+0.86}_{-0.77}$} & {4.69 $^{+0.22}_{-0.19}$} & {0.1596 $^{+0.0029}_{-0.0029}$} & 
          {0.15715 $^{+0.00251}_{-0.00254}$} & {5934.792530 $^{+0.000214}_{-0.000213}$} & {0.42 $^{+0.31}_{-0.31}$}  & {0.56} & {1.76} & {1123}\\\smallskip
       $H$  & {82.83 $^{+1.19}_{-1.04}$} & {5.00 $^{+0.31}_{-0.27}$} & {0.1533 $^{+0.0024}_{-0.0025}$} & 
          {0.15396 $^{+0.00245}_{-0.00248}$} & {5934.792508 $^{+0.000241}_{-0.000238}$} & {0.39 $^{+0.35}_{-0.34}$}  & {1.00} & {2.00} & 1234\\\smallskip
       $K$  & {82.42 $^{+0.84}_{-0.71}$} & {4.92 $^{+0.22}_{-0.19}$} & {0.1549 $^{+0.0022}_{-0.0022}$} & 
          {0.15465 $^{+0.00181}_{-0.00184}$} & {5934.792226 $^{+0.000233}_{-0.000232}$} & {-0.02 $^{+0.34}_{-0.33}$}  & {0.64} & {1.52} & {1270}\\
       \cline{2-10\smallskip}\smallskip
       (Weighted) & {82.61 $^{+0.20}_{-0.19}$} & {4.954 $^{+0.051}_{-0.048}$} & {0.15714 $^{+0.00064}_{-0.00065}$} 
          & \multicolumn{1}{c}{...} & {5934.792178 $^{+0.000040}_{-0.000040}$} & {-0.09 $^{+0.06}_{-0.06}$} & ... & ...\\
       \hline
       \multicolumn{9}{c}{{\it This work: Method 2 -- global joint analysis with common radius}}\dotfill\\\smallskip
        & {82.69 $^{+0.18}_{-0.18}$} & {4.979 $^{+0.048}_{-0.048}$} & ... & {0.15739 $^{+0.00065}_{-0.00064}$}
        & {5934.792190 $^{+0.000043}_{-0.000043}$} & -0.07 $^{+0.06}_{-0.06}$ & ... & ...\\
       \hline
       \multicolumn{9}{c}{{\it This work: Method 3 -- global joint analysis with wavelength-dependent radius} {\bf (adopted as the final results})}\dotfill\\\smallskip
        & {82.64 $^{+0.20}_{-0.19}$} & {4.967 $^{+0.051}_{-0.050}$} & ... & {0.15743 $^{+0.00041}_{-0.00041}$}
        \tablefootmark{a} & {5934.792193 $^{+0.000043}_{-0.000043}$} & {-0.07 $^{+0.06}_{-0.06}$} & ... & ...\\
       \hline
       \multicolumn{9}{c}{{\it \citet{gillon2012}}}\dotfill\\
        & 82.33 $^{+0.20}_{-0.20}$ & 4.918 $^{+0.053}_{-0.051}$ & ... & 0.15945 $^{+0.00076}_{-0.00077}$ &  
        5726.54336 $^{+0.00012}_{-0.00012}$ & 0.80 $^{+0.17}_{-0.17}$ & ... & ...\smallskip\\
       \hline
     \end{tabular}
     \tablefoot{\small
       \tablefoottext{a}{In the global joint analysis with wavelength-dependent radius (i.e. Method 3), $R_{\rm{p}}/R_{\star}$ is 
                         allowed to vary from filter to filter similar to Method 1. They are listed together with 
                         Method 1 for direct comparison. The final $R_{\rm{p}}/R_{\star}$ from Method 3 is a weighted 
                         mean of seven filters.}
       \tablefoottext{b}{$T_{\mathrm{mid}}$ is a jump parameter in the MCMC analysis.}
       \tablefoottext{c}{The O--C values are calculated from comparison to: 
                         ${ T(N)=2455934.792239(40)+N\times0.81347437(13)}$, where $N=0$ for this work 
                         and $N=-256$ for \citet{gillon2012}.}
       \tablefoottext{d}{{ The last three columns} in this table are calculated from the individual analysis, 
                         which are different from those in Table~\ref{tab:log}.}
     }
   \end{table*}

  Our seven-band transit light curves were obtained simultaneously, and covered a large 
  wavelength range from the optical to the NIR. While most inferred physical properties should 
  be principally the same in different bands, the probed apparent planetary "radius" 
  could be wavelength-dependent. Therefore we adopted three methods of MCMC analysis to determine 
  the system parameters and to investigate their deviations: 
   \begin{itemize} \itemsep1pt \parskip0pt \parsep0pt
     \item[--] {\it Method 1: individual analysis.} The seven transit light curves were 
          fitted individually, in which $i$, $a/R_{\star}$, $R_{\rm{p}}/R_{\star}$ and 
          $T_{\rm{mid}}$ were obtained for each light curve.
     \item[--] {\it Method 2: global joint analysis with common radius.} The seven transit 
          light curves were jointly fitted, in which all the light curves shared the same 
          $i$, $a/R_{\star}$, $R_{\rm{p}}/R_{\star}$ and $T_{\rm{mid}}$.
     \item[--] {\it Method 3: global joint analysis with wavelength-dependent radius.} The seven transit 
          light curves were jointly fitted, in which all the light curves shared the same 
          $i$, $a/R_{\star}$ and $T_{\rm{mid}}$, while $R_{\rm{p}}/R_{\star}$ was allowed 
          to vary from filter to filter.
   \end{itemize}
  
  In Method 1, we tried to characterize the light curves individually, so that the variation 
  of derived parameters from filter to filter can be investigated. Overall, the derived 
  parameters are consistent with each other. The standard deviations of individual $i$ and 
  $a/R_{\star}$ are well below their average uncertainties, while that of $R_{\rm{p}}/R_{\star}$ 
  and $T_{\rm{mid}}$ are less than twice of their average uncertainties. Strong correlation 
  between $i$ and $a/R_{\star}$ can be seen (see the data points with error bars in 
  Fig.~\ref{fig:tranpdf}), and both of them are correlated with $R_{\rm{p}}/R_{\star}$. 
  Only $T_{\rm{mid}}$ is independent of other parameters. Most red noise factors come from the 
  time-averaging method, except in the $g'$- and $H$-bands. These factors are very close 
  to unity in the optical, indicating that the level of red noise is low. It could also be a 
  result of relatively long exposure time (90 sec) that makes the cadence not frequent enough 
  to reflect the feature of red noise. The level of time-correlated noise in the NIR seems 
  obvious even with visual inspection, especially the dip between +50 and +80 minutes as shown 
  in Fig.~\ref{fig:tran}, thus resulting in larger rescaling factors. We estimated the standard 
  deviation of best-fit residuals per 2 minute intervals to show the quality of the light curve, 
  which can be directly compared to \citet{gillon2012}. We achieved a scatter of {$\sim$0.05--0.08\%} 
  in the optical, and {$\sim$0.11--0.13\%} in the NIR per 2 minute intervals, which 
  correspond to a range of {$\sim$2.1--4.5 and $\sim$6.2--8.7} times the photon noise limit, 
  respectively. { For comparison, \citet{gillon2012} reached accuracies $\sim$0.11--0.15\% in 
  their optical transit light curves using 60-cm and 1.2-m telescopes, and $\sim$0.04--0.05\% in 
  their NIR occultation light curves using an 8.2-m telescope.}
  
  In Methods 2 and 3, we tried to obtain transit parameters from a global view, in which 
  we can avoid introducing extra variation from the correlation between parameter pairs (e.g. 
  $i$ and $a/R_{\star}$). The differences between Methods 2 and 3 depend mainly on whether the 
  apparent planetary radius is wavelength-dependent. We also calculated the weighted average 
  of $R_{\rm{p}}/R_{\star}$ for Method 3 in the same way as we did in Method 1. 
  
  Results from these three methods agree very well with each other within their error bars. 
  Since physically the probed atmospheric depth could be different 
  in different bands due to either clouds/hazes or atomic/molecular absorption, we decide to report 
  results from Method 3 as our final results. Note that the red noise factors are larger in 
  the cases of global joint analysis, which is expected since time-correlated noise is coupled 
  with light curves and individual analysis has more flexibility in the modeling. The red noise 
  factors of Method 3 are shown in Table~\ref{tab:log} while those of Method 1 are shown in 
  Table~\ref{tab:lcpar}.

  \subsection{Fitting of occultation light curves}\label{sec303}
  
  \begin{figure}
     \centering
     \includegraphics[width=0.48\hsize]{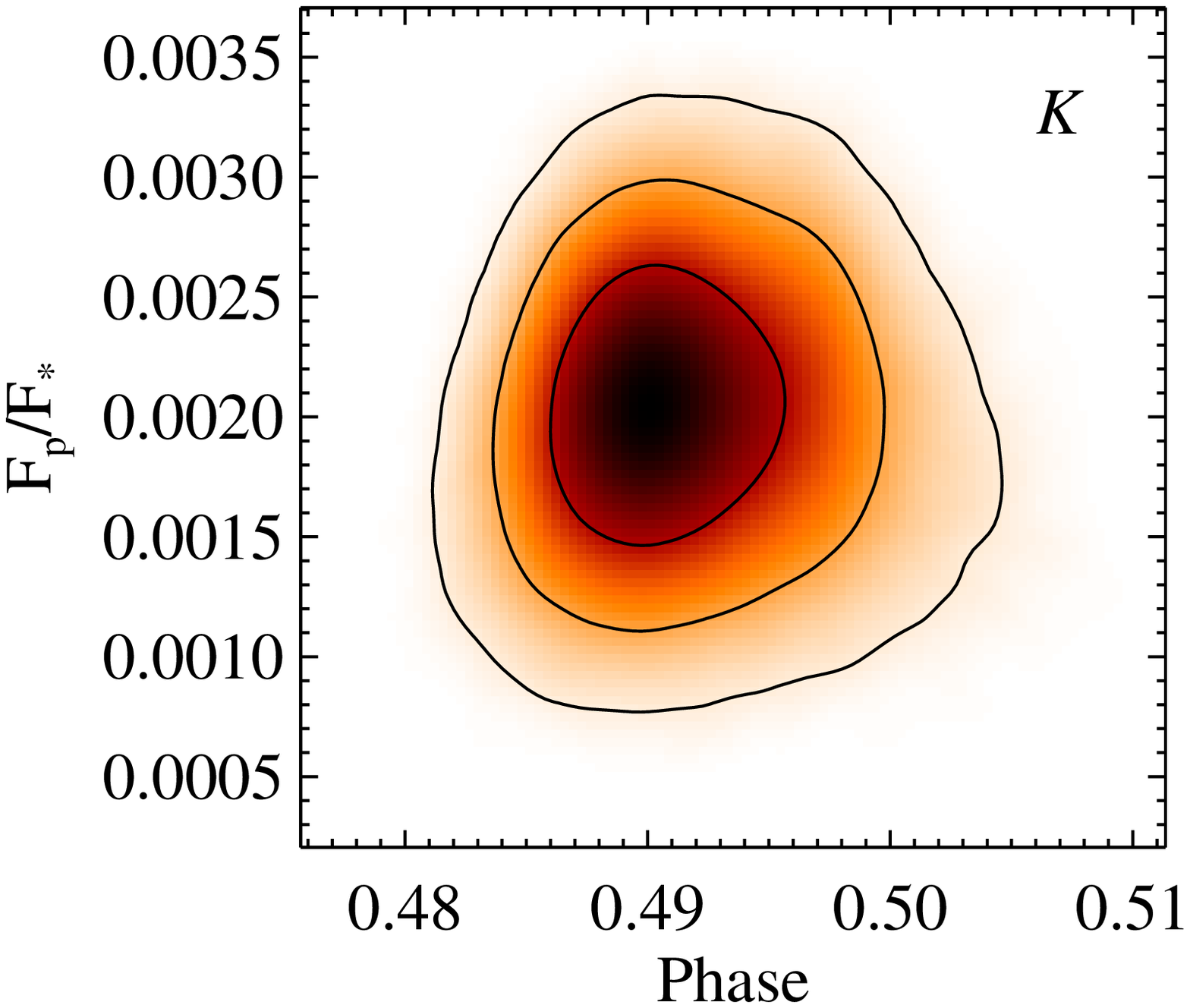}
     \includegraphics[width=0.48\hsize]{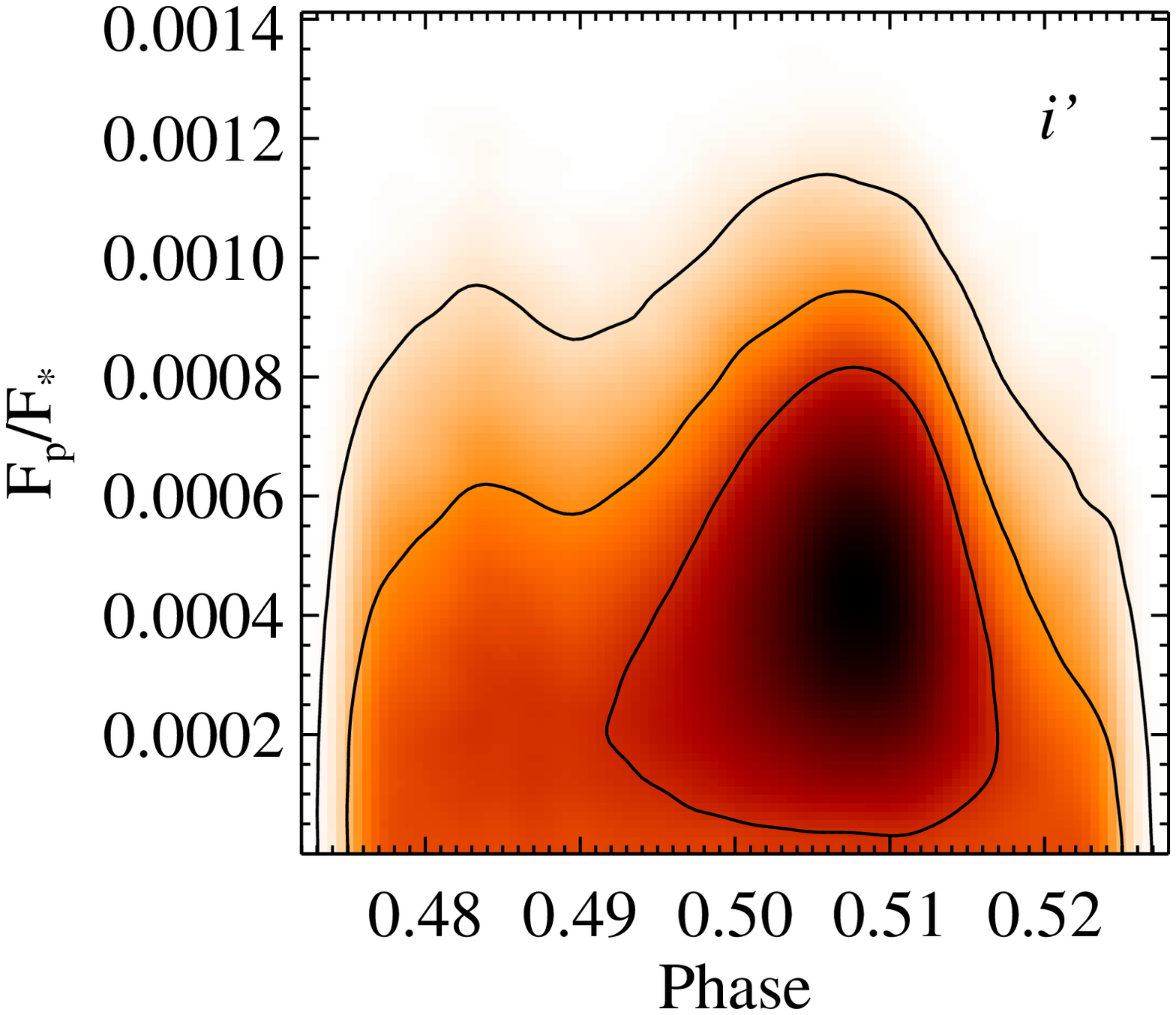}
     \caption{Correlation between mid-occultation time (converted to phase for display purpose) 
              and flux ratio for the $K$- and $i'$-bands derived from the MCMC analysis. Three 
              contour levels indicate the 68.3\% (1-$\sigma$), 95.4\% (2-$\sigma$) and 99.7\% 
              (3-$\sigma$) confidence levels, respectively.}
     \label{fig:occpdf}
   \end{figure}
  
  { Given the modest light-curve quality, we fitted the seven occultation 
  light curves by fixing the mid-occultation time $T_{\rm{occ}}$ and common 
  transit parameters at known values, and only setting $F_{\rm{p}}/F_{\star}$ 
  as the free parameter. Values of $a/R_{\star}$, $R_{\rm{p}}/R_{\star}$ and 
  $i$ were obtained from the global joint analysis on our transit light curves 
  (Method 3). $T_{\rm{occ}}$ was calculated from $T_{\rm{occ}}=T_0+(N+\phi)P$, 
  where $N=67$ and $\phi$ = 0.5002 $\pm$ 0.0004. The mid-occultation phase 
  $\phi$ was taken from \citet{2013arXiv1302.7003B}, which was precisely 
  determined by the joint analysis of {\it Spitzer} 3.6~$\mu$m and 
  4.5~$\mu$m occultation light curves. We only detected or marginally detected 
  the occultation signals in the $K$- and $i'$-bands. }
  
  { The flux ratios detected in the $K$- and $i'$-bands are 0.197 $\pm$ 
  0.042\% and 0.037 $^{+0.023}_{-0.021}$\%, respectively. Our $i'$-band detection is 
  marginal with 1.8-$\sigma$ significance. Our $K$-band detection agrees within 1-$\sigma$ with the 
  value of 0.156 $\pm$ 0.014\% obtained with VLT/HAWK-I in the 2.09\,$\mu$m narrow band \citep{gillon2012}, 
  and agrees well with the value of 0.194 $\pm$ 0.029\% obtained with CFHT/WIRCam in the $K_S$-band 
  \citep{2013ApJ...770...70W}. The time-correlated noise level is low for both $K$- and 
  $i'$-bands as shown in Fig.~\ref{fig:rmsbin}. We achieved a scatter of 1425~ppm and 604~ppm 
  per 2 minute intervals for the $K$- and $i'$-bands, with corresponding estimated photon noise 
  limits of {2.1$\times$10$^{-4}$} and 1.9$\times$10$^{-4}$, respectively.}
  
  { We also tried freely fitting $T_{\rm{occ}}$ in addition to $F_{\rm{p}}/F_{\star}$ for 
  the $K$- and $i'$-bands, so as to test the dependence of flux ratio on mid-occultation time. 
  This resulted in $F_{\rm{p}}/F_{\star}$ = 0.203 $\pm$ 0.036\% and 
  $F_{\rm{p}}/F_{\star}$ = 0.037 $^{+0.023}_{-0.021}$\%, respectively. As shown in Fig.~\ref{fig:occpdf}, 
  while the $i'$-band $T_{\rm{occ}}$ is not well constrained, the $K$-band $T_{\rm{occ}}$ occurs 
  on an offset phase around $\phi=0.4907^{+0.0036}_{-0.0027}$ (i.e. an offset of -10.9 minutes to 
  $\phi=0.5002$). Since both {\it Spitzer} and high-precision ground-based occultation 
  observations have ruled out an eccentric orbit for \object{WASP-43b} 
  \citep{gillon2012,2013arXiv1302.7003B,2013ApJ...770...70W}, we speculate the offset of our 
  $T_{\rm{occ}}$ arising from contamination of instrumental systematics. However, we note that 
  the occultation depths do not strongly depend on phase. The measured depths remain nearly constant 
  within 1-$\sigma$ when $T_{\rm{occ}}$ varies from phase 0.48 to 0.51. }

  \subsection{Fitting of mid-transit times}\label{sec304}
   \begin{table}[h!]
     \caption{Transit timing data used in this work}
     \label{tab:timing}
     \scriptsize
     \centering
     \begin{tabular}{llrc}
       \hline\hline
       Epoch & $T_{\mathrm{tran}}$$-$2\,450\,000 & O--C\tablefootmark{a} & Reference\tablefootmark{b}\\
             & [BJD$_{\mathrm{TDB}}$] & [min] &\\
       \hline
       ...\tablefootmark{c} & 5528.868227$^{+0.000078}_{-0.000078}$ & ... & \citet{gillon2012}\smallskip\\
       -2  &  5933.16500$^{+0.00035 }_{-0.00033}$ & { -0.42$^{+0.50}_{-0.48}$}  & 1\smallskip\\\smallskip 
       0   & { 5934.791932$^{+0.000114}_{-0.000116}$} & { -0.44$^{+0.16}_{-0.17}$} & This work ($g'$)\\\smallskip  
       0   & { 5934.792132$^{+0.000072}_{-0.000070}$} & { -0.15$^{+0.10}_{-0.10}$} & This work ($r'$)\\\smallskip  
       0   &  5934.792231$^{+0.000085}_{-0.000088}$ & { -0.01$^{+0.12}_{-0.13}$} & This work ($i'$)\\\smallskip  
       0   & { 5934.792407$^{+0.000107}_{-0.000110}$} & { 0.24$^{+0.15}_{-0.16} $} & This work ($z'$)\\\smallskip   
       0   & { 5934.792551$^{+0.000247}_{-0.000257}$} & { 0.45$^{+0.36}_{-0.37} $} & This work ($J$)\\\smallskip   
       0   & { 5934.792508$^{+0.000344}_{-0.000341}$} & { 0.39$^{+0.49}_{-0.49} $} & This work ($H$)\\\smallskip   
       0   & { 5934.792228$^{+0.000229}_{-0.000230}$} & { -0.02$^{+0.33}_{-0.33} $} & This work ($K$)\\\smallskip    
       6   &  5939.67471$^{+0.00082 }_{-0.00080}$ & { 2.34$^{+1.18}_{-1.15} $} & 2\\\smallskip 
       7   &  5940.48713$^{+0.00070 }_{-0.00061}$ & { 0.82$^{+1.01}_{-0.88} $}  & 3\\\smallskip     
       12  &  5944.55336$^{+0.00192 }_{-0.00202}$ & { -0.82$^{+2.76}_{-2.91}$}  & 4\\\smallskip  
       55  &  5979.53402$^{+0.00093 }_{-0.00090}$ & { 1.00$^{+1.34}_{-1.30} $}  & 5\\\smallskip   
       55  &  5979.53352$^{+0.00055 }_{-0.00058}$ & { 0.28$^{+0.79}_{-0.84} $}  & 6\\\smallskip   
       61  &  5984.41454$^{+0.00106 }_{-0.00088}$ & { 0.53$^{+1.53}_{-1.27} $}  & 7\\\smallskip   
       61  &  5984.41457$^{+0.00151 }_{-0.00147}$ & { 0.57$^{+2.17}_{-2.12} $}  & 8\\\smallskip   
       61  &  5984.41489$^{+0.00057 }_{-0.00059}$ & { 1.03$^{+0.82}_{-0.85} $}  & 9\\\smallskip   
       61  &  5984.41443$^{+0.00115 }_{-0.00109}$ & { 0.37$^{+1.66}_{-1.57} $}  & 8\\\smallskip   
       77  &  5997.43061$^{+0.00078 }_{-0.00082}$ & { 1.22$^{+1.12}_{-1.18} $}  & 10\\\smallskip   
       77  &  5997.43015$^{+0.00086 }_{-0.00090}$ & { 0.55$^{+1.24}_{-1.30} $}  & 5\\\smallskip   
       77  &  5997.42948$^{+0.00105 }_{-0.00104}$ & { -0.41$^{+1.51}_{-1.50}$}  & 6\\\smallskip  
       77  &  5997.42957$^{+0.00105 }_{-0.00111}$ & { -0.28$^{+1.51}_{-1.60}$}  & 6\\\smallskip  
       82  &  6001.49842$^{+0.00183 }_{-0.00247}$ & { 1.85$^{+2.64}_{-3.56} $}  & 11\\\smallskip   
       82  &  6001.49691$^{+0.00062 }_{-0.00070}$ & { -0.33$^{+0.89}_{-1.01}$}  & 12\\\smallskip    
       88  &  6006.37954$^{+0.00140 }_{-0.00149}$ & { 2.24$^{+2.02}_{-2.15} $}  & 8\\\smallskip   
       99  &  6015.32727$^{+0.00089 }_{-0.00101}$ & { 1.54$^{+1.28}_{-1.45} $}  & 13\\\smallskip   
       124 &  6035.66483$^{+0.00098 }_{-0.00104}$ & { 2.55$^{+1.41}_{-1.50} $}  & 14\\\smallskip   
       429 &  6283.77162$^{+0.00253 }_{-0.00224}$ & { -1.61$^{+3.64}_{-3.23}$}  & 15\\\smallskip  
       435 &  6288.65335$^{+0.00135 }_{-0.00211}$ & { -0.34$^{+1.94}_{-3.04}$}  & 16\\\smallskip  
       482 &  6326.88704$^{+0.00088 }_{-0.00094}$ & { 0.23$^{+1.27}_{-1.35} $}  & 17\\\smallskip   
       484 &  6328.51416$^{+0.00080 }_{-0.00075}$ & { 0.47$^{+1.15}_{-1.08} $}  & 18\\\smallskip    
       493 &  6335.83414$^{+0.00218 }_{-0.00190}$ & { -1.38$^{+3.14}_{-2.74}$}  & 17\\\smallskip  
       533 &  6368.37418$^{+0.00122 }_{-0.00134}$ & { 0.15$^{+1.76}_{-1.93}$}  & 19\\\smallskip  
       557 &  6387.89805$^{+0.00062 }_{-0.00065}$ & { 0.85$^{+1.89}_{-0.94}$}  & 17\\\smallskip  
       622 &  6440.77245$^{+0.00044 }_{-0.00042}$ & { -1.22$^{+0.63}_{-0.60}$}  & 17\\  
       \hline
     \end{tabular}
     \tablefoot{\small
          \tablefoottext{a}{The O--C values are calculated from comparison to linear 
                            regression: ${ T(N)=2455934.792239(40)+N\times0.81347437(13)}$.}
          \tablefoottext{b}{The numbers in the reference column indicate the source 
                            authors of the TRESCA project, which provides data at the website: 
                            http://var2.astro.cz/EN/index.php. TRESCA timings listed 
                            here are derived from a homogeneous fit using our own codes.}
          \tablefoottext{c}{This ephemeris entry refers to the 23 transits listed in the 
                            Table 6 of \citet{gillon2012}.}
     }
     \tablebib{\small
     (1) Starr, P.; (2) Naves, R.; (3) Ayiomamitis, A.; 
     (4) Ren\'{e}, R.; (5) Nicolas, E.; (6) Gonzalez, J.; 
     (7) Horta, F. G.; (8) Lomoz, F.; (9) Martineli, F.; 
     (10) Garcia, F.; (11) Carre\~{n}o, A.; (12) Schteinman, G. M.; 
     (13) Zibar, M.; (14) Hall, G.; (15) Chapman, A \& D\'{i}az, N. D.;
     (16) Lopesino, J.; (17) Evans, P.; (18) Haro J. L.;
     (19) B\"{u}chner, A.; 
     }
   \end{table}
  We performed another individual analysis on the seven transit light curves in order 
  to investigate the central transit times under the control of common system parameters. 
  We adopted $i$, $a/R_{\star}$ and $R_{\rm{p}}/R_{\star}$ determined in the global 
  joint analysis (Method 3), and imported them as Gaussian priors in this analysis. 
  Only $T_{\rm{mid}}$ was allowed to freely float throughout this modeling. This results 
  in seven measurements on the same epoch. As shown in Fig.~\ref{fig:timing}, the seven 
  central times are not exactly the same. The standard deviation of their best-fit values 
  is $\sim$1.3 times of their mean uncertainties. Compared to the uncertainty-weighted 
  average, the $r'$-, $i'$-, $H$-, $K$-bands deviate less than 1$\sigma$, the $z'$-, 
  $J$-bands deviate less than 2$\sigma$, while the $g'$-band deviates $\sim$2.2$\sigma$. 
  We repeated this analysis with different baseline models to examine whether this deviation 
  was caused by detrending. However, even when we forced all the light curves to have the 
  same simplest model (normalization factor plus a slope, i.e. $c_0+c_1t$) we could see a 
  similar trend. Therefore, if this deviation comes from imperfect light curve shape, we 
  cannot correct it by introducing baseline correction. Another possibility is that the 
  uncertainties of these central times are still underestimated. Considering that all the 
  timings deviate from the average by much less than 2$\sigma$ (except $g'$), this deviation 
  is not significant.
  
  We also re-analyzed the amateur transit light curves obtained from the TRESCA 
  Project\footnote{TRESCA is an acronym from words TRansiting ExoplanetS and CAndidates, 
  see http://var2.astro.cz/EN/tresca}\citep{2010NewA...15..297P}, which will be used in 
  the investigation of transit timing variations (TTVs) effect in Sect.~\ref{sec401}. 
  We discarded 15 out of 42 available light curves to date (August 2013), because of either 
  partial transits or very poor data quality (with a scatter in best-fit residuals larger 
  than 1\%) or asymmetric transit shape caused by strong visible time-correlated noise. 
  Since most of these data lack the instrumental parameters while some of them even do 
  not have measurement uncertainties, the instrumental systematics that might affect the 
  transit shape can hardly be corrected. In order to determine the central transit times 
  more robustly, we replaced our fitting statistic (i.e. $\chi^2$) with the wavelet-based 
  likelihood function proposed by \citet{2009ApJ...704...51C}, in which we assume the 
  time-correlated noise to vary with a power spectral density of 1/$f$ at frequency $f$. 
  The jump rule is the same, except that the likelihood increases with higher probability. 
  We converted the time stamps of TRESCA light curves into BJD$_{\mathrm{TDB}}$ and adopted 
  $i$, $a/R_{\star}$ and $R_{\rm{p}}/R_{\star}$ as Gaussian priors from our analysis described 
  in Sect.~\ref{sec302}. This resulted in uncertainties for the TRESCA central transit 
  times that are $\sim$20--270\% (on average 80\%) larger than those originally listed in 
  the TRESCA website. The newly determined central transit times for the TRESCA light curves 
  are shown in Table~\ref{tab:timing}. As a comparison, the timings of our seven transit 
  light curves would have been enlarged $\sim$10\% using this wavelet-based analysis. However, 
  in order to have a direct comparison with \citet{gillon2012}, for our transit light curves, 
  we choose the timings from the $\chi^2$-based analysis as the final values.
   
\section{Results and discussion}\label{sec4}
  
  \subsection{Period determination}\label{sec401}
   \begin{figure}
     \centering
     \includegraphics[width=\hsize]{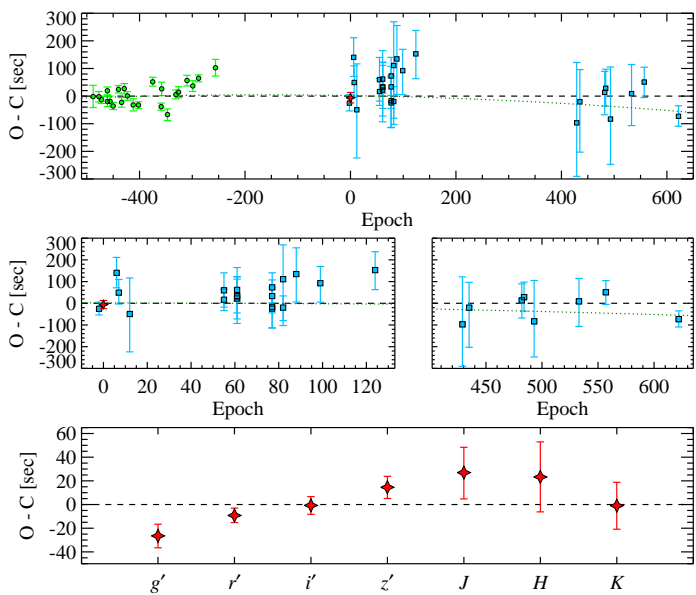}
     \caption{Transit timing O--C residuals for \object{WASP-43b}. In the {\it top panel}, 
              green circles show timings from \citet{gillon2012} (with epoch less than -200), 
              blue squares show timings from TRESCA (with epoch larger than -10, see 
              Sect.~\ref{sec304} for the re-analysis process), while red star shows our 
              weighted average timing value (at epoch 0), the error bar of which indicates 
              the standard deviation of our seven filters. {\it Middle panel} shows a zoom-in 
              view of the transit timings obtained after epoch -10. {\it Bottom panel} shows 
              our seven-filter timings. Dashed line refers to the linear regression, while 
              dotted line represents a quadratic fit.}
     \label{fig:timing}
   \end{figure}
  In addition to our seven timing measurements on one transit and 27 re-analyzed timings from 
  TRESCA, we also collected the 23 timings listed in Table 6 of \citet{gillon2012}. We fitted 
  a linear function to these 57 timings by minimizing $\chi^2$:
  \begin{equation}
    T_c(E)=T_c(0)+EP
  \end{equation}
  in which $E$ is the epoch, and the epoch number of our observation was set as 0. From the fitting 
  we obtained a new transit ephemeris of {$T_c(0)$ = 2455934.792239 $\pm$ 0.000040}\,($\rm{BJD_{TDB}}$) and 
  an improved orbital period of {$P$ = 0.81347437 $\pm$ 0.00000013}\,days. This new period is 4 times more 
  precise than that of \citet{gillon2012}. We show the O--C diagrams in Fig.~\ref{fig:timing}, and list all 
  the newly determined timing data in Table~\ref{tab:timing}. 
  
  This fit has a {$\chi^2=139.68$} with 55 degrees of freedom (DOF), resulting in a reduced chi-square 
  $\chi^2_{\nu}=2.59$. This indicates a poor fit by linear function. However, the standard deviation 
  for the best-fit residuals is 52.6~s, while the median uncertainty for individual timings is 30.2~s. 
  From this view, the significance of TTV is low. We examined all the timings, and found that four 
  epochs ($-$375, $-$347, $-$288 and $-$256) deviating more than 3$\sigma$ level lead to this large 
  $\chi^2_{\nu}$, all of which come from \citet{gillon2012}. Discarding these four epochs would result 
  in a smaller value {$\chi^2_{\nu}$=1.62}. We also note that, in the experiment of Sect.~\ref{sec304}, the 
  wavelet-based analysis produces on average 10\% larger uncertainties than our time-averaging and 
  prayer-bead methods. Considering that \citet{gillon2012} performed a similar analysis to ours which 
  is based on the timing-averaging method only, it is likely that this poor fit arises from underestimated 
  timing uncertainties. 
  
  \citet{2013arXiv1302.7003B} tried to fit all the available timings and to estimate the decay rate 
  using an quadratic ephemeris model \citep{2010ApJ...721.1829A}:
  \begin{equation}
    T_c(E)=T_c(0)+EP+\delta P*E(E-1)/2
  \end{equation}
  They obtained a result strongly favoring this model over the linear one. Since their results were 
  based on the original TRESCA timings with underestimated uncertainties, we decided to repeat the 
  same quadratic fit, which had a result of {$T_c(0)$ = 2455934.792262 $\pm$ 0.000041, 
  $P$ = 0.81347399 $\pm$ 0.00000022}~days and $\delta P$ = ${(-2.2\pm1.1)\times10^{-9}}$~days~orbit$^{-2}$. This 
  $\delta P$ translates into a $\dot{P}$ = 0.09 $\pm$ 0.04~s~year$^{-1}$, which is 7 times smaller than 
  that of \citet{2013arXiv1302.7003B}. This quadratic fit has a { BIC of 147.36}, which is almost the 
  same as that of the linear fit ({ BIC=147.69}). Therefore we can conclude that a quadratic model does 
  not improve the fit of ephemeris. 
  
  As a conclusion, we prefer no evidence for significant TTV within current dataset. The current 
  TTV rms of 52.6~s roughly corresponds to a timing deviation caused by a perturber with a mass of 
  2.17~$M_{\oplus}$ in the 2:1 resonance according to the relationship derived by \citet{2005MNRAS.359..567A}.

  \subsection{Physical parameters}\label{sec402}
   \begin{table*}
     \caption{System parameters of the \object{WASP-43} system}
     \label{tab:param}
     \small
     \centering
     \begin{tabular}{p{6cm}llc}
       \hline\hline
       Parameter & Symbol & Value & Note\\
       \hline
       \multicolumn{4}{c}{\it Transit Parameters}\\
       \hline
       Orbital period [days]\dotfill                   & $P$                & {0.81347437 $\pm$ 0.00000013}   & A\\
       Mid-transit time [$\mathrm{BJD_{TDB}}$]\dotfill & $T_{0}$            & {2455934.792239 $\pm$ 0.000040} & A\\
       Planet/star radius ratio\dotfill                & $R_{\rm{p}}/R_{\star}$   & {0.15743 $\pm$ 0.00041}         & B\\
       Orbital inclination [deg]\dotfill               & $i$                & {82.64 $\pm$ 0.19}              & B\\
       Scaled semi-major axis\dotfill                  & $a/R_{\star}$      & {4.967 $\pm$ 0.050}             & B\\
       Transit duration [days]\dotfill                 & $T_{14}$           & {0.05115 $\pm$ 0.00022}         & D\\
       Ingress/egress duration [days]\dotfill          & $T_{12}$=$T_{34}$  & {0.01103 $\pm$ 0.00025}         & D\\
       Transit impact factor\dotfill                   & $b=a\cos i/R_{\star}$    & {0.636 $^{+0.010}_{-0.011}$}     & D\\
       \hline
       \multicolumn{4}{c}{\it Occultation Parameters}\\
       \hline
       Mid-occultation time [$\mathrm{BJD_{TDB}}$]\dotfill & $T_{\mathrm{occ}}$ & {2455989.6943 $^{+0.0029}_{-0.0022}$} & C\\
       Planet-to-star flux ratio in $K$-band [\%]\dotfill  & $F_{\mathrm{p}}/F_{\star}$ & {0.197 $\pm$ 0.042} & C\\
       Planet-to-star flux ratio in $i'$-band [\%]\dotfill & $F_{\mathrm{p}}/F_{\star}$ & {0.037 $^{+0.023}_{-0.021}$} & C\\
       \hline
       \multicolumn{4}{c}{\it Other Orbital Parameters}\\
       \hline 
       Orbital eccentricity\dotfill                    & $e$        & 0. (fixed)                 & G\\
       Argument of periastron [$^{\circ}$]\dotfill     & $\omega$   & 0. (fixed)                 & G\\
       Orbital semi-major axis [AU]\dotfill            & $a$        & {0.01524 $\pm$ 0.00025} & F\\
       Roche limit [AU]\dotfill                        & $a_R$      & {0.00749 $\pm$ 0.00012} & F\\
       \hline
       \multicolumn{4}{c}{{\it Stellar Parameters}}\\
       \hline
       Mass [$M_{\sun}$]\dotfill               & $M_{\star}$            & {0.713 $^{+0.018}_{-0.021}$} & E\\
       Radius [$R_{\sun}$]\dotfill             & $R_{\star}$            & {0.660 $^{+0.008}_{-0.009}$} & F\\
       Density [$\rho_{\sun}$]\dotfill         & $\rho_{\star}$         & {2.482 $^{+0.077}_{-0.073}$} & D\\
       Surface gravity [cgs]\dotfill           & $\log g_{\star}$       & {4.652 $\pm$ 0.006} & F\\
       Effective temperature [K]\dotfill       & $T_{\rm{eff}}$         & {4536 $^{+98}_{-85}$}        & E\\
       Metallicity [dex]\dotfill               & $[\rm{Fe/H}]$          & 0.01 $^{+0.10}_{-0.09}$    & E\\
       Age [Gyr]\dotfill                       & $t_{\rm{age}}$         & {4.4 $^{+3.7}_{-2.4}$}       & E\\
       \hline
       \multicolumn{4}{c}{{\it Planetary Parameters}}\\
       \hline
       Mass [$M_{\rm{Jup}}$]\dotfill            & $M_{\rm{p}}$                  & {2.029 $^{+0.035}_{-0.040}$} & F\\
       Radius [$R_{\rm{Jup}}$]\dotfill          & $R_{\rm{p}}$                  & {1.034 $\pm$ 0.014}          & F\\
       Density [g\,cm$^{-3}$]\dotfill           & $\rho_{\rm{p}}$               & {2.434 $^{+0.067}_{-0.065}$} & F\\
       Surface gravity [cgs]\dotfill            & $\log g_{\rm{p}}$             & {3.692 $\pm$ 0.009}          & D\\
       Equilibrium temperature ($A_B$=0, $f$=1/4) [K] \dotfill      & $T_{\rm{eq}}$          & {1439 $^{+31}_{-28}$}        & D\\
       Equilibrium temperature ($A_B$=0, $f$=1/2) [K] \dotfill      & $T_{\rm{eq}}$          & {1712 $^{+37}_{-33}$}        & D\\
       Equilibrium temperature ($A_B$=0, $f$=2/3) [K] \dotfill      & $T_{\rm{eq}}$          & {1839 $^{+40}_{-36}$}        & D\\
       Brightness temperature in $K$-band [K]\dotfill               & $T_{\mathrm{B},K}$     & {1878 $^{+108}_{-116}$}      & D\\
       Brightness temperature in $i'$-band [K]\dotfill              & $T_{\mathrm{B},i'}$     & {2225 $^{+139}_{-225}$}      & D\\
       Incident flux [10$^9$~erg\,s$^{-1}$\,cm$^{-2}$]\dotfill      & $\langle F\rangle$     & {0.973 $^{+0.088}_{-0.073}$} & D\\
       Safronov Number\dotfill                  & $\Theta$                      & {0.0873 $^{+0.016}_{-0.015}$} & D\\
       \hline
     \end{tabular}
     \tablefoot{
       A: Determined from linear regression of timing data listed in Table~\ref{tab:timing}; 
       B: Determined from our global joint analysis of the seven-band transit light curves (for detail, see
          Table~\ref{tab:lcpar});
       C: Determined from our analysis of the $K$- or $i'$-band occultation light curves;
       D: Derived from the PDF of group B or C using theoretical formulae;
       E: Derived from the evolutionary tracks analysis using PDF from group B;
       F: Derived from the PDF of group B and E using theoretical formulae;
       G: Both are fixed as 0 in our analysis of transit light curves.
     }
   \end{table*}
  \begin{figure}
     \centering
     \includegraphics[width=\hsize]{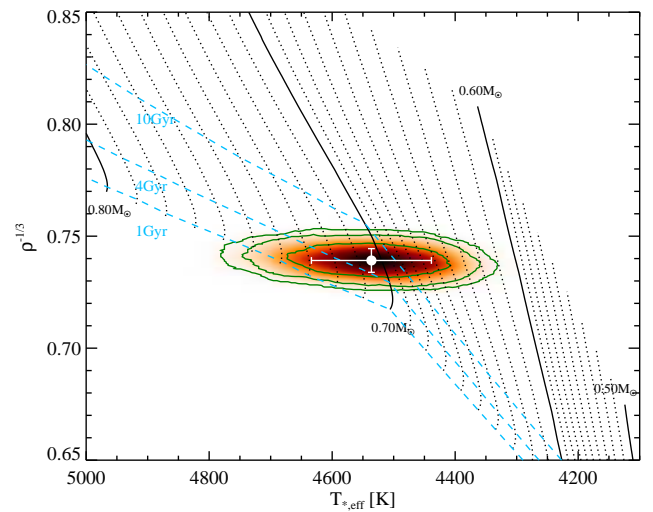}
     \caption{Stellar evolutionary tracks obtained from \citet{2012A&A...541A..41M} to derive the stellar 
              mass and age for \object{WASP-43}. Solid lines labeled with numbers show the basic tracks 
              at solar abundance, while dotted lines show the interpolated tracks. Dashed lines 
              indicate the isochrones. Contours overlaid on the distribution density map show the 
              posterior distribution of links in our MCMC process, in which those with different 
              metallicities are also included.}
     \label{fig:isochrone}
   \end{figure}
  
  In order to determine the physical parameters for the \object{WASP-43} system, 
  we made use of the output PDFs from our MCMC analysis on the transit light curves, 
  stellar evolutionary tracks, as well as the RV semi-amplitude. We first 
  derived a set of parameters that simply depended on the light curve, such as 
  transit duration, ingress/egress duration and mean stellar density, using the 
  formulae from \citet{2003ApJ...585.1038S}. By adopting the RV semi-amplitude 
  value (in the form of $K_2=K\sqrt{1-e^2}P^{1/3}$) from \citet{gillon2012}, 
  we were able to derive the planetary surface gravity \citep{2007MNRAS.379L..11S}. 
  
  We then determined the stellar mass and age by interpolating in the stellar 
  evolutionary tracks using the MCMC process. Since photometric measurements are 
  independent of spectoscopic measurements that provide us stellar effective 
  temperature $T_{\rm{eff}}$ and metallicity [Fe/H], it is a good complement 
  to the determination of stellar mass. To have a direct comparison with 
  \citet{gillon2012}, we adopted the Geneva stellar evolutionary tracks 
  \citep{2012A&A...541A..41M}, and interpolated them into finer grids. The 
  mass interpolation was performed in the [$\rho_*^{-1/3}$, $T_{\rm{eff}}$] 
  plane by MCMC, in which, $T_{\rm{eff}}$ and [Fe/H] were obtained from 
  \citet{gillon2012}, and were randomly generated from Gaussian distributions, 
  while $\rho_*^{-1/3}$ was taken from the $a/R_{\star}$ chains derived in our 
  previous global joint analysis. At each step, the link [$\rho_*^{-1/3}$, 
  $T_{\rm{eff}}$] was put onto the tracks according to [Fe/H]. Interpolation 
  was bilinearly performed among mass tracks and isochrones. Links that were 
  off tracks or older than 12Gyr were discarded. After this MCMC process, we 
  derived a stellar mass of {0.713 $^{+0.018}_{-0.021}~M_{\sun}$} and a poorly 
  constrained age of {4.4 $^{+3.7}_{-2.4}$\,Gyr} for WASP-43. The posterior 
  distributions of $T_{\rm{eff}}$ ({4536 $^{+98}_{-85}$\,K}) and [Fe/H] 
  (0.01 $^{+0.10}_{-0.09}$) remain similar to input, the best-fit values of 
  which only shift slightly since our $a/R_{\star}$ is slightly different from 
  \citet{gillon2012}. Fig.~\ref{fig:isochrone} shows the posterior distribution 
  of this MCMC process. With this refined stellar mass value, new stellar radius, 
  orbital semi-major axis, and planetary radius were derived. Finally, the 
  planetary mass was derived by solving the Equation 25 of \citet{2010arXiv1001.2010W}.
  
  We obtained a mass of {2.029 $^{+0.035}_{-0.040}$}~$M_{\rm{Jup}}$ and a radius of 
  {1.034 $\pm$ 0.014}~$R_{\rm{Jup}}$ for \object{WASP-43b}, which correspond to a bulk 
  density of {2.434 $^{+0.067}_{-0.065}$~g~cm$^{-3}$}. This puts \object{WASP-43b} in 
  the top list of dense hot Jupiters. \citet{2007ApJ...659.1661F} calculated groups 
  of giant planet thermal evolution models, in which the mass-radius relationships 
  are parameterized by age, core mass and effective orbital distance. Assuming that 
  \object{WASP-43b} is a hypothetical planet moving around the Sun which receives 
  the same flux as the actual planet, the effective orbital distance is calculated 
  as $a_{\oplus}=a(L_{\star}/L_{\sun})^{-1/2}=0.0376$~AU. According to the theoretical 
  prediction, only models with an old age could explain current data. After interpolating 
  in the theoretical models with a solar composition at 4.5~Gyr, we obtained theoretical 
  planetary radii of 1.13~$R_{\rm{Jup}}$, 1.07~$R_{\rm{Jup}}$, and 1.01~$R_{\rm{Jup}}$ 
  { in the cases of core free, a 50~$M_{\oplus}$ core, and a 100~$M_{\oplus}$ core, 
  respectively. Therefore, the current mass and radius} for \object{WASP-43b} indicate a massive 
  core inside this planet, and favor an old age that is consistent with the derived age from 
  stellar evolutionary tracks.

  \subsection{Atmospheric properties}\label{sec403}
  One of our primary goals observing both transit and occultation events is to probe 
  the atmosphere of \object{WASP-43b} in a complementary way, since the atmosphere of both 
  dayside and terminator region have been observed through these two events. 
  
  For the transit light curves, we did the re-analysis by fixing the values of $i$, 
  $a/R_{\star}$ and $T_{\rm{mid}}$ to those from the final global joint analysis  
  (Method 3), while letting the limb-darkening coefficients float under the control 
  of Gaussian priors. We derived the transit depth for each band and list them in 
  Table~\ref{tab:spec}. The resulting $R_{\mathrm{p}}/R_{\star}$ are slightly more 
  precise than those determined in the global joint analysis, because the uncertainties 
  of $i$, $a/R_{\star}$ and $T_{\rm{mid}}$ were not propagated. { Because we seek 
  to derive the conditional distribution of transit depth in each band, the uncertainties 
  arising from common parameters are not taken into account in the derived transmission spectrum.}
  
  For the occultation light curves, we also placed 3-$\sigma$ upper limits on the 
  potential occultation depths in the other five bands in addition to the $K$- and 
  $i'$-band detections, which are listed in Table~\ref{tab:spec}.
  
   \begin{table}
     \caption{Transit and occultation depths for transmission and emission spectra}
     \label{tab:spec}
     \small
     \centering
     \begin{tabular}{ccc}
       \hline\hline
       Filter & $(R_{\mathrm{p}}/R_{\star})^2$ (\%)\tablefootmark{a} & $F_{\mathrm{p}}/F_{\star}$ (\%)\\
       \hline
       $g'$ & {2.481 $\pm$ 0.023} & $<$0.086\tablefootmark{b}\\
       $r'$ & {2.478 $\pm$ 0.016} & {$<$0.083}\tablefootmark{b}\\
       $i'$ & {2.511 $\pm$ 0.017} & {0.037 $^{+0.023}_{-0.021}$}\\
       $z'$ & {2.472 $\pm$ 0.023} & {$<$0.094}\tablefootmark{b}\\
       $J$ &  {2.469 $\pm$ 0.077} & {$<$0.186}\tablefootmark{b}\\
       $H$ &  {2.371 $\pm$ 0.075} & {$<$0.254}\tablefootmark{b}\\
       $K$ &  {2.392 $\pm$ 0.056} & {0.197 $\pm$ 0.042}\\
       \hline
     \end{tabular}
     \tablefoot{\small
       \tablefoottext{a}{In this fit, only $R_{\mathrm{p}}/R_{\star}$, $u_1$ (with prior) and $u_2$ (with prior) 
                         are allowed to vary, while others are fixed at the values from the global joint fit.}
       \tablefoottext{b}{3-$\sigma$ upper limit is placed when no detection.}
     }
   \end{table}
   \subsubsection{A broad-band transmission spectrum}\label{sec40301}
   \begin{figure}
     \centering
     \includegraphics[width=\hsize]{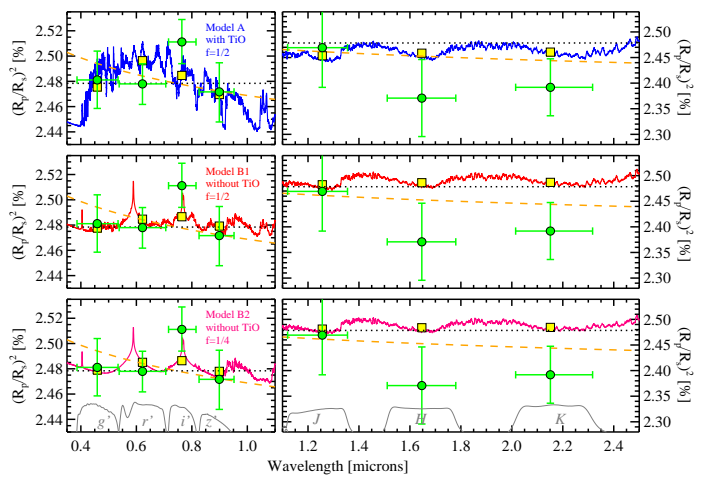}
     \caption{Model transmission spectra compared to observed transit depths which are 
              derived from a single transit. In all panels, green circles with error bars 
              show our measurements in the $g'r'i'z'JHK$ bands, of which the horizontal 
              error bars indicate the FWHM of each bandpass. Black dotted line shows a 
              constant value of $R_{\rm{p}}^2/R_{\star}^2=2.48\%$. Orange dashed line 
              shows a Rayleigh scattering spectrum which is caused by a high-altitude 
              haze. Three fiducial model spectra are shown for comparison, and their 
              broad-band integrated values are displayed in yellow squares. The optical 
              and NIR data are shown in separate panels with different scales for clarity. 
              {\it Top panels} show a model with TiO/VO as the dominant opacity sources 
              and a dayside-wide averaged ($f=1/2$) pressure-temperature (P-T) profile. 
              {\it Middle} and {\it bottom panels} show two models with Na and K as the 
              dominant opacity sources. While the middle-panel model has a dayside-wide 
              averaged P-T profile, the bottom-panel model has a planet-wide averaged 
              P-T profile ($f=1/4$). See Sect.~\ref{sec40301} for detailed discussion. }
     \label{fig:transpec}
   \end{figure}
   We created a broad-band transmission spectrum by putting all the seven-band transit 
   depths together with respect to wavelength. By fitting this observed "spectrum" to 
   a flat straight line (see the black dotted lines in Fig.~\ref{fig:transpec}), we 
   obtain a constant transit depth of $R_{\rm{p}}^2/R_{\star}^2=2.48\%$ with {$\chi^2$=7.95} 
   (6 DOF), which indicates good agreement between the "spectrum" and a featureless line. 
   { The largest deviation comes from the $i'$-band (1.6$\sigma$), the $H$-band 
   (1.5$\sigma$) and the $K$-band (1.6$\sigma$). We confirm that the deviation at these 
   three bands does not arise from different detrending models for different bands}, 
   as this deviation shape still exists when forcing all seven light curves have the 
   same baseline function. A flat featureless spectrum could indicate an atmosphere 
   covered with high-altitude optically thick clouds. 
   
   The standard deviation of best-fit transit depth values is 0.051\%. Although the 
   insufficient precision ({$\sim$0.02\% and $\sim$0.07\%} for the optical and NIR bands, 
   respectively) prevents this variation far from being significant, we still try to 
   compare our "spectrum" to theoretical models, so as to examine whether a clear 
   atmosphere could apply. Detailed atmospheric modeling is beyond the scope of this 
   paper. Instead, we choose to generate fiducial atmospheric models based on the physical 
   and orbital parameters of the \object{WASP-43} system. The model atmosphere is computed 
   using the approach described in \citet{2005ApJ...627L..69F,2008ApJ...678.1419F}, which 
   uses the equilibrium chemistry mixing ratios from \citet{2002Icar..155..393L,
   2006asup.book....1L} and \citet{2009arXiv0910.0811L}, and adopts the opacity database 
   from \citet{2008ApJS..174..504F}. The formation of clouds or hazes is not included. 
   Three model transmission spectra (see the three panels in Fig.~\ref{fig:transpec}) 
   are calculated following the method of \citet{2010ApJ...709.1396F}, depending on how 
   the atmospheric pressure-temperature (P-T) profile is dealt with and what optical 
   opacity sources dominate. Model A has the P-T profile averaged over only the dayside 
   (i.e. heat redistribution factor $f$=1/2), the dominant opacities of which come from 
   gaseous TiO/VO. In contrast, model B1/B2 has gaseous Na and K as the dominant opacity 
   sources. While the P-T profile of B1 is similar to A, that of B2 is planet-wide averaged 
   (i.e. $f$=1/4). Broad-band model values are integrated from the synthetic transmission 
   spectra over each bandpass of GROND. 
   
   We fit the observed "spectrum" to the three model spectra by shifting the models 
   up-and-down, as the planetary base radius is unknown. The models are allowed to 
   have different base radii. We find that they fit the observed "spectrum" equally 
   well. The resulting {$\chi^2$ (6 DOF) are 6.69, 7.62, and 7.36} for models A, B1, 
   and B2, respectively. Fitting to fiducial atmospheric models only marginally 
   improves from that to a flat line, and the dominant atmospheric opacity sources 
   cannot be discerned. 
   
   We note that the best-fit transit depth values are in a decreasing trend from the 
   $g'$-band to the $K$-band, indicating that Rayleigh scattering could exist in the 
   upper atmosphere. Rayleigh scattering signature has been observed in the transmission 
   spectrum of \object{HD 189733b} obtained by several space-borne instruments 
   \citep[e.g.][]{2008A&A...481L..83L,2008MNRAS.385..109P,2011MNRAS.416.1443S,
   2012MNRAS.422..753G,2013MNRAS.432.2917P}. A recent work by \citet{2013MNRAS.432.2917P} 
   summarized that \object{HD 189733b} has a high-altitude haze of condensate grains 
   extending over at least five scale heights, which results in a Rayleigh scattering 
   slope from UV into NIR. We follow the approach of \citet[][see their Equation 
   1]{2008A&A...481L..83L} to construct a Rayleigh scattering spectrum for \object{WASP-43b}, 
   in which MgSiO$_3$ (with a size of $\sim$0.01~$\mu$m) is assumed as the haze condensate. 
   By shifting the Rayleigh scattering spectrum to match our data, we obtain { a good fit 
   with $\chi^2$=6.28} (6 DOF). Therefore, an atmosphere with high-altitude hazes could 
   also explain our observed "spectrum". The Rayleigh scattering spectrum is shown in 
   Fig.~\ref{fig:transpec} in orange dashed line.

   \subsubsection{Detection of dayside flux in the $K$- and $i'$-bands}\label{sec40302}
   \begin{figure}
     \centering
     \includegraphics[width=\hsize]{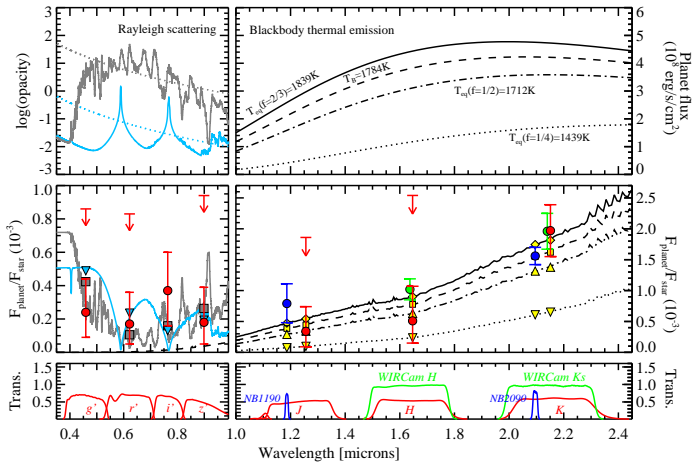}
     \caption{Dayside model spectra compared to the observed occultation depths.  
              In the {\it top panels}, the left one displays relative contributions 
              of absorption (solid curves) and Rayleigh scattering (dotted lines). 
              Two models of different opacity sources (TiO/VO in grey v.s. Na/K in 
              skyblue) are shown. The top-right panel displays blackbody planetary 
              dayside thermal emission spectra, with temperatures of { 1784~K, 1439~K, 
              1712~K, and 1839~K, corresponding to the 2.09~$\mu$m brightness temperature} 
              $T_B$ and equilibrium temperatures $T_{\rm{eq}}$ of three different heat 
              redistribution factors, respectively. {\it Middle panels} show the 
              planet-star flux ratios calculated based on {\it top panels}. Our 
              seven-band measurements are shown in red circles with error bars or 
              arrows (3$\sigma$ upper limit). Another two narrow-band measurements 
              from VLT/HAWK-I \citep{gillon2012} and two broad-band measurements from 
              CFHT/WIRCam \citep{2013ApJ...770...70W} are shown in blue and green 
              circles, respectively. Bandpass-integrated model values are plotted 
              in symbols without error bars. {\it Bottom panels} show the filter 
              transmission curves. See detailed discussion in Sect.~\ref{sec40302}.
              }
     \label{fig:occspec}
   \end{figure}
   We detected a flux ratio of {0.197 $\pm$ 0.042\%} in the GROND $K$-band, and {0.037 $^{+0.023}_{-0.021}$\%} 
   in the GROND $i'$-band. Before comparing them to theoretical models, we first estimate their 
   brightness temperatures. The planetary dayside thermal emission is simplified as blackbody, 
   while the stellar flux is interpolated from the Kurucz stellar models \citep{1979ApJS...40....1K} 
   with the values of $T_{\rm{eff}}$, $\log g$ and [Fe/H] derived from this work. The $K$-band 
   occultation detection is translated into a brightness temperature of {1878 $^{+108}_{-116}$}~K, 
   while the $i'$-band detection corresponds to {2225 $^{+139}_{-225}$}~K. 
   
   { With the equation $T_{\mathrm{eq}}=T_{\mathrm{eff,*}}\sqrt{R_{\star}/a}[f(1-A_B)]^{1/4}$ 
   \citep[where 1/4$\le$$f$$\le$2/3;][]{2011ApJ...729...54C}, we can infer the heat redistribution factor based on brightness 
   temperature. However, the brightness temperatures at both $K$- and $i'$-bands exceed the 
   maximum allowed equilibrium temperature 1839~K (when $f=2/3$ and $A_B=0$, i.e. heat is 
   re-emitted instantly without redistribution). Given the large uncertainties of brightness 
   temperature, our $K$-band still allows a heat redistribution factor of $f\ge0.56$ and 
   a bond albedo of $A_B\le0.16$ at 1-$\sigma$ level, indicative of very poor heat redistribution 
   efficiency. For the $i'$-band, the brightness temperature exceeds the maximum temperature at 
   1.7$\sigma$ level.} In contrast, an isothermal blackbody emission could provide a maximum 
   flux ratio of 0.006\% in the $i'$-band. 
   
   Since our $K$-band detection of thermal emission puts no more constraint than the VLT 2.09~$\mu$m 
   narrow-band detection, detailed atmospheric modeling is again beyond the scope of this paper. Given 
   that ground-available near-infrared bands contain only weak molecular features \citep{2012ApJ...758...36M} 
   and current data (including {\it Warm Spitzer}) of \object{WASP-43b} could not constrain the chemical 
   composition \citep{2013arXiv1302.7003B}, we decide to only investigate the atmospheric thermal emission 
   with simplified isothermal blackbody models. As shown in the right panels of  Fig.~\ref{fig:occspec}, 
   four blackbody models are displayed, with temperatures of four cases: { brightness temperature 
   of the 2.09~$\mu$m narrow band and equilibrium temperatures for $f$=1/4, $f$=1/2, and $f$=2/3. The 
   ground based detections are, within uncertainties, all consistent with the maximum-temperature spectrum. }
   The new analysis is hence consistent with previous studies \citep{gillon2012,2013arXiv1302.7003B,2013ApJ...770...70W}.
   
   However, our measurements in the optical cannot be explained by dayside thermal emission alone under 
   the simple blackbody assumption. Planets, however, do not necessarily radiate as black bodies 
   in these optical bands \citep[e.g.][]{2010A&A...513A..76S}. If we consider 
   the optical detection to be reflected light, we can calculate the geometric albedo using the 
   relationship $A_g=(F_{\mathrm{p}}/F_{\star})/(R_{\mathrm{p}}/a)^2$. The $i'$-band detection corresponds 
   to ${ A_g=0.37\pm0.22}$, while {3$\sigma$ upper limits of 0.86, 0.83, 0.94} are placed on the $g'$-, $r'$-, 
   $z'$-bands where no occultation is detected. { These values would change to ${ A_g=0.31\pm0.22}$, 
   and $A_g<$ 0.85, 0.81, 0.78, respectively, if the contamination by thermal emission is corrected for 
   assuming no heat redistribution ($f=2/3$).} Although the geometric albedo of most hot Jupiters obtained so 
   far seems to be low \citep[e.g.][]{2008ApJ...689.1345R,2011MNRAS.417L..88K,2011ApJS..197...11D}, which 
   is in accordance with cloud-free models of hot Jupiter atmospheres \citep[e.g.][]{2003ApJ...588.1121S,
   2008ApJ...682.1277B}, a high geometric albedo has been detected for some planets, e.g. \object{Kepler-7b} 
   \citep[0.32 $\pm$ 0.03,][]{2011ApJ...735L..12D} and \object{HD 189733b} 
   \citep[0.40 $\pm$ 0.12,][]{2013ApJ...772L..16E}. Our $i'$-band geometric albedo is comparable to these 
   high geometric albedos, but with large uncertainties. 
   
   We further tentatively investigate a Rayleigh scattering atmosphere for \object{WASP-43b} following 
   the approach of \citet{2013ApJ...772L..16E} to create toy models, in which thermal emission is ignored. 
   Depending on at what altitude the clouds or hazes become optically thick, we combine the atomic and molecular 
   absorption and Rayleigh scattering profiles to simulate the atmosphere. In the left panels of Fig.~\ref{fig:occspec}, 
   we show reflected light spectra calculated from Rayleigh scattering profiles mixed with two absorption profiles, 
   that are TiO/VO dominated or Na/K dominated which have been used for models A and B2 as described in 
   Sect.\ref{sec40301}. Basically the $i'$-band flux ratio can be well explained by reflective atmosphere 
   without taking other optical measurements into account, as the upper limits of the other three bands put 
   no meaningful constraints. If we assume the $g'$-band flux ratio 0.024 $^{+0.020}_{-0.015}$\% as a marginal 
   detection like the $i'$-band, all the optical measurements can still be explained by a reflective atmosphere, 
   but with a lower fraction of Rayleigh scattering. 
      
   An atmosphere with reflective clouds/hazes for \object{WASP-43b} is also allowed by our broad-band 
   transmission spectrum as discussed in Sect.~\ref{sec40301}. However, clouds/hazes should not be 
   highly reflective considering that the bond albedo is relatively low according to the $K$-band 
   thermal detection. Also, we cannot rule out the possibility of the $i'$-band detection being a 
   false positive. More observations in the optical are required to validate these measurements. 
   On the other hand, our $K$-band thermal detection confirms an irradiated atmosphere with poor 
   heat redistribution. Further investigation of chemical composition requires high-precision 
   spectroscopy in the NIR.

\section{Conclusions}\label{sec5}

   We observed one transit and one occultation event of \object{WASP-43b} 
   using the GROND instrument on the MPG/ESO 2.2-meter telescope. From 
   the simultaneously acquired $g'$, $r'$, $i'$, $z'$, $J$, $H$, $K$ 
   transit light curves, we have independently derived the planetary 
   system parameters, which have the same precision as that of 
   \citet{gillon2012} but have slightly different best-fit values. With 
   the newly derived mass {2.029 $^{+0.035}_{-0.040}$}~$M_{\rm{Jup}}$ and 
   radius {1.034 $\pm$ 0.014}~$R_{\rm{Jup}}$, we confirm that 
   \object{WASP-43b} is a relatively dense hot Jupiter with a massive 
   internal core. After collecting timings reported by \citet{gillon2012} 
   and the TRESCA timings that have been reanalyzed by the wavelet-based 
   method, we have derived a new ephemeris {$T_0$ = 2455934.792239 $\pm$ 
   0.000040}~days and an improved period {$P$ = 0.81347437 $\pm$ 
   0.00000013}~days. No significant TTV signal has been detected. 
   
   We have performed a tentative analysis on the wavelength dependent 
   transit depths. No significant variation in transit depths has been 
   found{, with the largest deviation of 1.6$\sigma$, 1.5$\sigma$, 
   and 1.6$\sigma$ in the $i'$-, $H$-, and $K$-bands, respectively.} Our broad-band 
   transmission spectrum can be explained by either a flat featureless 
   straight line that indicates thick clouds, atomic Na/K or molecular 
   TiO/VO imprinted spectra that indicate a cloud-free atmosphere, or 
   a Rayleigh scattering profile that indicates high-altitude hazes. 
   More high-precision observations in spectroscopic resolution are 
   required to discern the terminator atmosphere of \object{WASP-43b}. 
   
   We have detected the dayside thermal emission of \object{WASP-43b} 
   in the $K$-band with a flux ratio of {0.197 $\pm$ 0.042}\%, corresponding  
   to a brightness temperature of {1878 $^{+108}_{-116}$}~K. { Our $K$-band 
   detection is consistent with the \citet{gillon2012} detection in the 2.09~$\mu$m 
   narrow band and the \citet{2013ApJ...770...70W} detection in the $K_S$-band. }
   Thus we confirm that the dayside atmosphere is very inefficient in heat redistribution. 
   
   We have tentatively detected the dayside flux of \object{WASP-43b} 
   in the $i'$-band with a flux ratio of {0.037 $^{+0.023}_{-0.021}$\%}. Its 
   brightness temperature {2225 $^{+139}_{-225}$}~K is too high to be explained 
   based solely on the assumption of isothermal blackbody. We have preformed 
   tentative analysis involving Rayleigh scattering caused by reflective 
   hazes present in the upper atmosphere of the dayside. Firm conclusion 
   cannot be drawn based on current optical measurements. More high-precision 
   observations in the optical are required to validate the speculation of 
   reflective atmosphere.

\begin{acknowledgements}
      { We thank the anonymous referee for her/his careful reading 
      and helpful comments that improved the manuscript. }
      We acknowledge Markus Rabus and Timo Anguita for technical support of the 
      observations. G.C. acknowledges Chinese Academic of Sciences and Max Planck Society 
      for the support of doctoral training programme. H.W. acknowledges the support 
      by NSFC grants 11173060, 11127903, and 11233007. W.W. acknowledges the support 
      by NSFC grant 11203035. Part of the funding for GROND (both hardware as well as 
      personnel) was generously granted from the Leibniz-Prize to Prof. G. Hasinger 
      (DFG grant HA 1850/28-1).
\end{acknowledgements}


\clearpage

  \Online

  \begin{appendix}
  \section{Baseline correction of the light curves}\label{sec:ap01}
  We have described the light curve modeling process in Sect.~\ref{sec3}, in which a baseline correction is 
  applied at each MCMC step using the SVD algorithm. For light curves in every filter, we experimented a set 
  of baseline models composed of star position, seeing, airmass etc. The final choice was made by comparing 
  the BICs among different models. { In this experiment, when the fit is poor with large BIC, the transit 
  depths are very likely deviant from those determined by the chosen models, while the occultation 
  signals are always not detected. When the baseline models result in similar lowest BICs, the derived transit 
  depths (or flux ratios) become consistent with each other within their 1-$\sigma$ uncertainties. We also note that 
  the variation of best-fit transit depths (or flux ratios) caused by different baseline models is in accord 
  with the distribution derived by the chosen model.} Here we list the final models adopted for both transit 
  and occultation. Their coefficients are listed in Table \ref{tab:bf}. Those for occultation light curves 
  without detections are not shown for clarity. Note that the coefficients of transit light curves come from 
  the global joint analysis with wavelength-dependent radius (i.e. Method 3).
  \begin{eqnarray}
  B_{\mathrm{tran},g'} &=& c_0+c_1y+c_2t\\
  B_{\mathrm{tran},r'} &=& c_0+c_1x+c_2y+c_3xy+c_4y^2\\
  B_{\mathrm{tran},i'} &=& c_0+c_1t\\
  B_{\mathrm{tran},z'} &=& c_0+c_1x+c_2y\\
  B_{\mathrm{tran},J}  &=& c_0+c_1y+c_2t+c_3z\\
  B_{\mathrm{tran},H}  &=& c_0+c_1x+c_2y+c_3xy+c_4x^2\\
  B_{\mathrm{tran},K}  &=& c_0+c_1x+c_2y+c_3xy+c_4x^2+c_5y^2+c_6s\\
  B_{\mathrm{occ},i'}  &=& c_0+c_1s_y+c_2t+c_3t^2+c_4z\\
  B_{\mathrm{occ},K}   &=& c_0+c_1x+c_2s
  \end{eqnarray}

   \begin{table}
     \caption{Derived coefficients of the selected baseline models for both transit and occultation}
     \label{tab:bf}
     \small
     \centering
     \begin{tabular}{crrrrrrr}
       \hline\hline
       Coeff. & $g'$ & $r'$ & $i'$ & $z'$ & $J$ & $H$ & $K$\\
       \hline
      \multicolumn{8}{c}{{\it 2012-01-08 transit}}\dotfill\\
       $c_0$ & 0.999464(74) & 1.000102(74) & 0.999987(58)
            & 0.999845(97) & 1.00123(51) & 1.00125(31)
            & 1.00350(33)\\
       $c_1$ & 0.0001930(99) & 0.0000547(96) & 0.01840(28)
            & -0.000074(12) & 0.00030(82) & 0.000487(94)
            & 0.00442(41)\\
       $c_2$ & -0.03627(45) & 0.000019(29) & ...
            & -0.0001682(29) & -0.197(23) & 0.00224(13)
            & 0.00492(40)\\
       $c_3$ & ... & 0.0000838(54) & ...
            & ... & -0.0639(39) & -0.000788(34)
            & 0.00746(38)\\
       $c_4$ & ... & -0.000170(187) & ...
            & ... & ... & -0.00152(18)
            & 0.003036(92)\\
       $c_5$ & ... & ... & ...
            & ... & ... & ...
            & 0.00232(15)\\
       $c_6$ & ... & ... & ...
            & ... & ... & ...
            & 0.00065(18)\\
       \hline
       \multicolumn{8}{c}{{\it 2012-03-03 occultation}}\dotfill\\
       $c_0$ & ... & ... & 0.99665$^{+(73)}_{-(66)}$
            & ... & ... & ...
            & 1.00203(11)\\
       $c_1$ & ... & ... & 0.000194$^{+(33)}_{-(30)}$
            & ... & ... & ...
            & 0.001140(18)\\
       $c_2$ & ... & ... & 0.162$^{+(24)}_{-(27)}$
            & ... & ... & ...
            & 0.00571(22)\\
       $c_3$ & ... & ... & 4.66$^{+(74)}_{-(83)}$
            & ... & ... & ...
            & ...\\
       $c_4$ & ... & ... & -0.235$^{+(39)}_{-(35)}$
            & ... & ... & ...
            & ...\\
       \hline
     \end{tabular}
     \tablefoot{\small
          The two bracketed digits in super-/sub-script show the 68.3\% uncertainties of corresponding best-fit values. 
          They should be compared to the last two digits of the best-fit values. 
     }
   \end{table}

  \end{appendix}

\end{document}